\documentclass[journal]{IEEEtran} 

\usepackage{epsfig}
\usepackage{calc}

\usepackage{amsmath}
\usepackage{amssymb}
\usepackage{amsfonts}
\usepackage{epsfig}

\newcommand{\Q}[2]{\ensuremath{Q_{#1}^{(#2)}}}

\newtheorem{thm}{Theorem}
\newtheorem{prop}[thm]{Proposition}
\newtheorem{lem}[thm]{Lemma}

\newtheorem{note}{Remark}
\newtheorem{ex}{{\em Example}}

\newcommand{\epf}{\hfill $\Box$}

\newcommand{\beq}{\begin{equation}}
\newcommand{\eeq}{\end{equation}}

\newcommand{\bea}{\begin{eqnarray}}
\newcommand{\eea}{\end{eqnarray}}
\newcommand{\bean}{\begin{eqnarray*}}
\newcommand{\eean}{\end{eqnarray*}}

\newcommand{\bit}{\begin{itemize}}
\newcommand{\eit}{\end{itemize}}

\def\Q{\mathbb{Q}}
\def\R{\mathbb{R}}

\def\o{\omega}
\def\oi{\omega^{-1}}

\title{Perfect Space-Time Codes with Minimum and Non-Minimum Delay for Any Number of
Antennas}

\author{Petros Elia, B. A. Sethuraman and P. Vijay Kumar
\thanks{Petros Elia and P. Vijay Kumar are with the Department of EE-Systems,
University of Southern California, Los Angeles, CA 90089 ({\tt
\{elia,vijayk\}@usc.edu}).  B. A. Sethuraman is with the
Department of Mathematics,  of the California State University
Northridge, CA 91330 ({\tt al.sethuraman@csun.edu}). This work was
carried out while P. Vijay Kumar was on a leave of absence at the
Indian Institute of Science Bangalore. }} 
\begin{document}

\maketitle\thispagestyle{empty}
\begin{abstract}
Perfect space-time codes were first introduced by Oggier et. al.
to be the space-time codes that have full rate, full
diversity-gain, non-vanishing determinant for increasing spectral
efficiency, uniform average transmitted energy per antenna and
good shaping of the constellation.  These defining conditions
jointly correspond to optimality with respect to the Zheng-Tse
D-MG tradeoff, independent of channel statistics, as well as to
near optimality in maximizing mutual information. All the above
traits endow the code with error performance that is currently
unmatched. Yet perfect space-time codes have been constructed only
for $2,3,4$ and $6$ transmit antennas.  We construct minimum and
non-minimum delay perfect codes for all channel dimensions.
\end{abstract}

\section{Introduction\label{s:IntroductionPerfectCodes}}
\subsection{Definition of Perfect Codes \label{ss:Definition_Perfect_Codes}}
In \cite[Definition 1]{PerfectCodes}, the concept of perfect codes
is introduced to describe the $n \times n$ space-time codes that
satisfy all of the following criteria:
\begin{itemize}
\item \emph{Full rate}.  This corresponds to the ability of the
code to transmit $n$ symbols per channel use from a
\emph{discrete} constellation such as the QAM or the HEX
constellation. \item \emph{Full diversity}.  This corresponds to
having all $\Delta X \Delta X^\dag$, $\Delta X $ a difference
matrix of the code, be non-singular. \item \emph{Non vanishing
determinant for increasing spectral efficiency}.  The determinant
of any difference matrix, prior to SNR normalization, is lower
bounded by a constant that is greater than zero and independent of
the spectral efficiency. \item \emph{Good shaping of the
constellation}. When the code is based on cyclic division algebras
(CDA) (\cite{SetRajSas,BelRek,PerfectCodes,EliRajPawKumLu}), the
condition requires that the signalling set, in the form of the
layer-by-layer vectorization of the code-matrices, be isomorphic
to $\text{QAM}^{n^2}$ or $\text{HEX}^{n^2}$, where the isomorphism
is given \emph{strictly by some unitary matrix}. \item
\emph{Uniform average transmitted energy per antenna}. The
condition requires that the expected value of the transmitted
power is the same for all antennas. In fact we will see that the
structure of the code will allow for equal average power across
the antennas as well as across time.
\end{itemize}

The term \emph{perfect}, coined in \cite{PerfectCodes}, was in
reference to the ability of the codes to satisfy all the above
criteria, as well as in reference to the codes having the best
observed performance.
\subsection{New results \label{ss:NewResults}}
\subsubsection{Additional properties satisfied by perfect codes}
In addition to the defining properties, perfect codes also satisfy
\begin{itemize}
\item \emph{Approximate universality}. This property was
introduced in \cite{TavVisUniversal_2005} to describe a code that
is D-MG optimal \cite{ZheTse} irrespective of channel statistics.
Such codes exhibit high-SNR error performance that is given by the
high-SNR approximation of the probability of outage of any given
channel, and allow for the probability of decoding error given no
outage to vanish faster than the probability of outage. Currently,
the only family of approximately universal codes is that of
cyclic-division algebras with non-vanishing determinant, the
generalization of which was presented in \cite{EliRajPawKumLu}.
Consequently perfect codes are also approximately universal. \item
\emph{D-MG optimality for any spacial correlation of the fading or
the additive noise}. This can be immediately concluded by
post-multiplying the received signal matrix with the covariance
matrix of the additive noise vectors and then using the
approximate universality property of the code over channels with
uncorrelated additive white noise. \item \emph{Residual
approximate universality}. Any truncated code version resulting
from deletion of rows maintains approximate universality over the
corresponding channel. \item \emph{Gaussian-like signalling.} This
is an empirical observation and it relates to a Gaussian-like
signalling set with a covariance matrix that tends to maximize
mutual information. \item \emph{Information losslessness}. This
property relates to having full rate as well as unitary linear
dispersion matrices \cite{HassibiLDCs,HassibiLDCs_conference}, and
guarantees that the mutual information is not reduced as a result
of the code's structure. \item \emph{Scalable sphere decoding
complexity}. This property guarantees that as the number of
receive antennas becomes smaller, the structure of the code allows
for substantial reductions in sphere decoding complexity without
essential loss in performance. For MISO channels, the structure of
the code will allow for reduction of sphere decoding complexity,
from $O(n^2)$ to $O(n)$.
\end{itemize}

\subsubsection{Summary of presented contributions}

In this paper we introduce explicit constructions of minimum-delay
perfect space-time codes for any number $n$ of transmit antennas
and any number $n_r$ of receive antennas.  Non-minimum delay
perfect codes are constructed for any delay $T$ that is a multiple
of $n$. We also generalize the defining conditions in
\cite{PerfectCodes} to be independent of the channel topology,
thus directly providing for joint maximization of the mutual
information and approximate universality. For channels with a
smaller number of receive antennas, this information theoretic
approach allows for the construction of efficient variants of
perfect codes which exhibit almost the same error performance as
standard perfect codes but with substantially reduced sphere
decoding complexity.

\section{Satisfying the Perfect-Code Conditions}
\subsection{The general CDA structure\label{ssec:first three conditions}}
Our constructions of the perfect $n\times n$ space-time block
codes will be based on cyclic division algebras.

As shown in other related works such as
\cite{SetRajSas,BelRek,PerfectCodes,EliRajPawKumLu}, the basic
elements of a CDA space-time code are the number fields
$\mathbb{F}, \mathbb{L}$, with $\mathbb{L}$ a finite, cyclic
Galois extension of $\mathbb{F}$ of degree $n$. For $\sigma$ being
the generator of the Galois group
$\mbox{Gal}(\mathbb{L}/\mathbb{F})$, we let $z$ be some symbol
that satisfies the relations
\begin{equation}\label{eq:commutativity identity for div alg} \ell z \ = \ z \sigma(\ell) \ \ \
\forall \ \ \ell \in \mathbb{L} \ \ \ \mbox{ and } \ \ \
z^n=\gamma
\end{equation} for some `non-norm' element $\gamma \in
\mathbb{F}^\divideontimes:=\mathbb{F} \setminus \{0\}$ such that
the smallest positive integer $t$ for which $\gamma^t$ is the
relative norm $N_{\mathbb{L}/\mathbb{F}}(u)$ of some element $u$
in $\mathbb{L}^\divideontimes$, is $n$. The cyclic division
algebra is then constructed as a right $\mathbb{L}$ space
\begin{equation}\label{eq:div alg as a vector space}D= \mathbb{L}
\oplus z \mathbb{L} \oplus \hdots \oplus
z^{n-1}\mathbb{L}.\end{equation} A space-time code $\mathcal{X}$
can be associated to $D$ by selecting the set of matrices
corresponding to the representation by left multiplication on $D$
of elements from a finite subset of $D$. For an arbitrary choice
of \textit{scaled} integral basis $\{\beta_i\}_{i=0}^{n-1}$ of
$\mathbb{L}$ over $\mathbb{F}$ (by a \textsl{scaled integral
basis} we mean a set of numbers obtained by multiplying all the
elements of an integral basis by the same nonzero real number),
the $n$-tuple $\{f_{i,j}\}_{i=0}^{n-1}$ maps to
\begin{equation}\label{eq:element of maximal field}\ell_j =
\sum_{i=0}^{n-1} f_{i,j} \beta_i,\in \mathbb{L}, \  \ \ \
f_{i,j}\in \mathcal{O}_\mathbb{F}
\end{equation} where $\mathcal{O}_\mathbb{F}$ is the ring of integers of $\mathbb{F}$.
Consequently, prior to SNR normalization, the code-matrix $X$
representing the division algebra element $x = \sum_{j=0}^{n-1}
z^j \ell_j, \ \ell_j\in \mathbb{L}$, is given by the defining
equations (\ref{eq:commutativity identity for div alg}-\ref{eq:div
alg as a vector space}) to be
\begin{eqnarray} \label{eq:CDA_ST_Codes1}
X=\begin{array}{|ccccc|}
\ell_0  &  \gamma \sigma(\ell_{n-1})  & \gamma \sigma^2(\ell_{n-2}) & \! \cdots  \!    &  \gamma \sigma^{n-1}(\ell_{1})  \\
\ell_1  &  \sigma(\ell_0)             & \gamma \sigma^2(\ell_{n-1}) & \!  \cdots  \!    &   \gamma \sigma^{n-1}(\ell_{2}) \\
\vdots  &                       &                          &           &   \vdots                     \\
\ell_{n-2} &  \sigma(\ell_{n-3})      & \sigma(\ell_{n-4})          & \!   \cdots  \!   & \gamma \sigma^{n-2}(\ell_{n-1})    \\
\ell_{n-1} &  \sigma(\ell_{n-2})      & \sigma(\ell_{n-3})          & \!   \cdots  \!   & \sigma^{n-1}(\ell_0)    \\
\end{array}
\end{eqnarray}
or equivalently
\begin{equation} \label{eq:distributed_CDA} X= \sum_{j=0}^{n-1}
\Gamma^j \biggl( diag \bigl( \underline{f}_{j}\cdot G \bigr)
\biggr)
\end{equation}
with ${ \underline{f}_j =[ f_{j,0} \  f_{j,1} \ \cdots \ f_{j,n-1}
]}$ where
 \begin{eqnarray}\label{eq:G_Matrix}G & = & \begin{array}{|ccc|}
\beta_0  &  \cdots &   \sigma^{n-1}(\beta_{0})\\
&  \vdots & \\
\beta_{n-1} & \cdots &
 \sigma^{n-1}(\beta_{n-1})\\
\end{array} \\[5pt] \label{eq:GammaMatrix}  \ \  \Gamma & = & \begin{array}{|ccccc|}
0 & 0 & \cdots  & 0 & \gamma\\
1 & 0 & \cdots  & 0 & 0\\
0 & 1 & \cdots  & 0 & 0\\
 & & \vdots & &\\
0 & 0 & \cdots  & 1 & 0\end{array} \ .
\end{eqnarray}
For this setup, it was shown in
\cite{BelRek}\cite{KirRaj}\cite{EliRajPawKumLu}, that if $\gamma$
is independent of SNR and the $f_{i,j}$ are from a discrete
constellation such as QAM, then the code achieves
\begin{enumerate}\item full diversity \item full-rate \item
non-vanishing determinant.\end{enumerate}

\subsubsection{Requirements for achieving equal power sharing and good constellation}
Based on the above setup, we now show that using a unit-magnitude,
algebraic, non-norm element $\gamma$, and a unitary $G$, renders
the code perfect.

Let $\mathcal{B} := \{\beta_0,\cdots,\beta_{n-1}\}$ be a basis for
$\mathbb{L}/\mathbb{F}$ such that the matrix
$$G(\mathcal{B}) = \begin{array}{|ccc|} \beta_0 & \cdots & \sigma^{n-1}(\beta_0) \\ & \vdots & \\\beta_{n-1} & \cdots &
\sigma^{n-1}(\beta_{n-1})\end{array}$$ is unitary.  Furthermore,
we assume that $|\gamma| = 1$.

Let $\mathbb{F} = \mathbb{Q}(\imath)$ and let the $\{f_{i,j} \}$
be restricted to belong to the $M^2$-QAM constellation:
$$f_{i,j}\in \mathcal{A}_{\text{QAM}} = \{a+\imath b \ |  \ -(M-1)\leq a,b \leq M-1, \ a,b  \ \text{both odd} \}. $$
For $$\underline{f} := [f_{0,0} \ f_{0,1} \ \cdots f_{0,n-1} \
f_{1,0} \cdots f_{1,n-1}\cdots f_{n-1,n-1}]$$ we denote the
code-matrix $X$ in (\ref{eq:CDA_ST_Codes1}) as $X(\underline{f})$
to emphasize that it is a function of the QAM vector
$\underline{f}$. If $g$ is any function of the $\{ f_{i,j} \}$, we
use $\mathbb{E}[g(\{f_{i,j}\})]$ to denote the average $\frac{1}{|
\mathcal{A}^{n^2}_{\text{QAM}}|}\sum\limits_{f_{i,j}\in
\mathcal{A}^{n^2}_{\text{QAM}}} g(\{f_{i,j}\}).$ Then for $k_0\in
\{ 0,1\}$, it is the case that
\begin{eqnarray*} \mathbb{E}[|\gamma^{k_0}\sigma^t(\ell_j)|^2] & = &
\mathbb{E}[(\sum_i
f_{i,j}\sigma^t(\beta_i))(\sum_k f^{*}_{k,j} \sigma^t(\beta^*_k))] \\
& = &
 \mathbb{E}[\sum_i |f_{i,j}|^2 | \sigma^t (\beta_i)|^2] \\
 & = & \sum_i \mathbb{E}[|f_{i,j}|^2 | \sigma^t (\beta_i)|^2 ]  \\
 & = & \mathbb{E}[|f_{i,j}|^2] \sum_i |\sigma^t(\beta_i)|^2 \\
 & = &
 \mathbb{E}[|f_{i,j}|^2].\end{eqnarray*}
It follows that the average energy per transmitted element of the
code matrix is the same.

With respect to the constellation shaping, let us now denote the
layer-by-layer vectorization of $X(\underline{f})$ as:
$$\text{vec}(X(\underline{f})) :=
\begin{array}{|cccc|}
I_{n\times n}             & 0_{n\times n} & \cdots & 0_{n\times n} \\
0_{n\times n} & \cdot\Gamma^{(1)}       & \cdots & 0_{n\times n} \\
              &               & \vdots              &  \\
0_{n\times n} & 0_{n\times n} &  \cdots & \cdot\Gamma^{(n-1)} \\
\end{array} \
\begin{array}{|c|}\ell_0 \\ \sigma(\ell_0) \\ \vdots \\ \sigma^{n-1}(\ell_0) \\ \vdots \\ \ell_{n-1} \\ \vdots \\ \sigma^{n-1}(\ell_{n-1})
\end{array}$$
where
\begin{eqnarray*} \Gamma^{(1)} & = & \begin{array}{|ccccc|}
        1 & 0 & .. & 0 & 0      \\
          &   & \vdots &   &        \\
        0 & 0 & .. & 1      & 0      \\
        0 & 0 & .. & 0      & \gamma \\ \end{array}, \
\Gamma^{(2)}  =  \begin{array}{|ccccc|}
        1 & 0 & .. & 0 & 0      \\
          &   & \vdots &   &        \\
        0 & 0 & .. & \gamma & 0      \\
        0 & 0 & .. & 0      & \gamma \\ \end{array} \\
                               & \vdots &         \\
\Gamma^{(n-1)} & = & \begin{array}{|ccccc|}
        1 & 0 & .. & 0 & 0      \\
        0 & \gamma & .. & 0 & 0      \\
          &   & \vdots &   &        \\
        0 & 0 & .. & \gamma & 0      \\
        0 & 0 & .. & 0 & \gamma \\ \end{array}.
\end{eqnarray*}
We observe that each vector resulting from the layer-by-layer
vectorization of any code-matrix, prior to SNR normalization, is
exactly the linear transformation of the $n^2$-tuple
$\underline{f}$ from $\mbox{QAM}^{n^2}$ or $\mbox{HEX}^{n^2}$, by
the unitary matrix
\begin{equation}\label{eq:vectorization matrix} R_v =
\begin{array}{|ccccc|}
G             & 0_{n\times n} & 0_{n\times n} & \cdots & 0_{n\times n} \\
0_{n\times n} & G\cdot\Gamma^{(1)}       & 0_{n\times n} & \cdots & 0_{n\times n} \\
0_{n\times n} & 0_{n\times n} & G\cdot\Gamma^{(2)}       & \cdots & 0_{n\times n} \\
              &               & \vdots        &        &  \\
0_{n\times n} & 0_{n\times n} & 0_{n\times n} & \cdots & G\cdot\Gamma^{(n-1)} \\
\end{array}.\end{equation}
As a result the signalling set prior to SNR normalization is from
the lattice
\begin{equation}\label{eq:signaling set lattice}\Lambda_\text{signal} = \{ \underline{f}R_v, \
\underline{f}\in \mbox{QAM}^{n^2} \}\end{equation} or
$\Lambda_\text{signal} = \{ \underline{f}R_v, \ \underline{f}\in
\mbox{HEX}^{n^2} \}$. Modifying $\underline{f}$ by horizontally
stacking its real and imaginary parts, $\underline{f}_{Re}$ and
$\underline{f}_{Im}$ respectively, and transforming the resulting
vector $\underline{f}^{'} = [\underline{f}_{Re} \ \
\underline{f}_{Im}] \in \mathbb{Z}^{2n^2}$ by
\begin{equation} \label{eq:vectorized_Real_Imag_Stacking Lattice Gen
Matrix} R^{'}_v = \begin{array}{|cc|} R_{v,Re}  & R_{v,Im} \\
-R_{v,Im} & R_{v,Re}
\end{array}
\end{equation}
we get the real and imaginary stacking of the signalling set to be
from
\begin{equation}\label{eq:signaling set
lattice real imag stacking} \Lambda^{'}_\text{signal} = \{
\underline{f}^{'}R^{'}_v, \ \underline{f}^{'}\in
\mathbb{Z}^{2n^2}\}, \ \ \ \ R^{'}_{v} R^{'T}_v = I_{2n^2\times
2n^2}
\end{equation}
exactly satisfying the good constellation shaping condition as
described in \cite[Definition 1]{PerfectCodes}.

Equivalently, for any set of $k_i\in \{0,1\}$,
$$\begin{array}{||c||} \gamma^{k_0}\ell_j\\ \gamma^{k_1}\sigma(\ell_j) \\ \vdots\\ \gamma^{k_{n-1}}\sigma^{n-1}(\ell_j)\end{array}^{ \ 2} = \begin{array}{||c||} \ell_j\\ \sigma(\ell_j) \\ \vdots\\ \sigma^{n-1}(\ell_j)\end{array}^{ \ 2} = \begin{array}{||c||} f_{0,j}\\ f_{1,j} \\ \vdots\\ f_{n-1,j}\end{array}^{ \ 2}$$
allowing for $$ Tr(X^\dag X) = \|\text{vec}(X(\underline{f}))\|^2
= \|\underline{f}\|^2$$ and making the collection
$$ \{ \text{vec}(X(\underline{f})) \ | \
\underline{f}\in \mathcal{A}^{n^2}_{\text{QAM}} \}$$ represent a
cubic constellation in $n^2$-dimensional complex space that is
isometric to $\mathcal{A}^{n^2}_{\text{QAM}}.$

We proceed to find a proper unit-magnitude non-norm element
$\gamma$, and then a proper unitary matrix $G$.

\subsection{Uniform average transmitted energy per antenna}
We now provide algebraic, unit-magnitude `non-norm' elements
$\gamma$ for suitable cyclic Galois extensions
$\mathbb{L}/\mathbb{F}$, independent of SNR. We will henceforth
denote the $l^{\text{th}}$ primitive root of unity as $\o_l$, i.e.
$\o_l = e^{2\pi\imath/l}$, and by $k^*$ we will denote the complex
conjugate of $k\in \mathbb{C}$. We directly state the construction
method for the different cases of interest, depending on whether
the base field is $\mathbb{F} = \mathbb{Q}(\imath)$ (QAM) or
$\mathbb{F} = \mathbb{Q}(\o_3)$ (HEX).

\begin{prop}\label{prop:nonNormUnitMagnQAM}
(\textit{Construction of the non-norm element for the QAM code})
Let $n = 2^{s}n_1 $ where $n_1$ is odd. Then there exists a prime
$p$ congruent to $1$ mod $n_1$. Furthermore, there exists a prime
$q$ that is congruent to $1$ mod $4$, as well as congruent to $5$
mod $2^{s+2}$, and which has order
$\text{ord}(q)|_{\mathbb{Z}_p^{\divideontimes}} = n_1$ and splits
in $\mathbb{Z}[\imath]$ as $q = \pi_1\pi_1^{*}$ for a suitable
prime $\pi_1 \in \mathbb{Z}[\imath]$. The fields
$\mathbb{Q}(\o_p)$ and $\mathbb{Q}(\imath)$ are linearly disjoint
over $\mathbb{Q}$. Let $\mathbb{K}$ be the unique subfield of
$\mathbb{Q}(\imath)(\o_p)$ of degree $n_1$ over
$\mathbb{Q}(\imath)$ and let $\mathbb{L} = \mathbb{K}\cdot
\mathbb{Q}(\o_{2^{s+2}})$. Then $\mathbb{L}$ is a cyclic extension
of $\mathbb{Q}(\imath)$, and the element
$$\gamma = \frac{\pi_1}{\pi_1^{*}}$$ is an (algebraic)
unit-magnitude element that is a non-norm element for the
extension $\mathbb{L}/\mathbb{Q}(\imath)$ and is independent of
SNR. (When $n_1 = 1$, we take $\mathbb{L} =
\mathbb{Q}(\o_{2^{s+2}})$ and $\gamma =
\displaystyle{\frac{1+2\imath}{1-2\imath}}$.)
 The $n\times n$ matrices arising from
equations (\ref{eq:distributed_CDA}), (\ref{eq:G_Matrix}) and
(\ref{eq:GammaMatrix}) with the above choice of
$\mathbb{F},\mathbb{L}$ and $\gamma$, and with any choice of
scaled integral basis $\{\beta_i\}_{i=0}^{n-1}$ hence yield a
full-diversity, full-rate code over QAM with non-vanishing
determinant satisfying the additional equal power sharing
constraint.
\end{prop}
\begin{proof} See Appendix \ref{sec:Appendix PerfectCodes NonNorm}
\end{proof}

\begin{prop}\label{prop:nonNormUnitMagnHEX}
(\textit{Construction of the non-norm element for the HEX code})
Let $n = 2^sn_1, \ s\in\{0,1\} $,  where $n_1$ is odd. Then there
exists a prime $p>3$ congruent to $1$ mod $n_1$. Furthermore,
there exists a prime $q$ that is congruent to $1$ mod $3$ and
which has order $\text{ord}(q)|_{\mathbb{Z}_p^{\divideontimes}} =
n_1$ and splits in $\mathbb{Z}[\o_3]$ as $q = \pi_1\pi_1^{*}$ for
a suitable prime $\pi_1 \in \mathbb{Z}[\o_3]$.  If $s=1$ then $q$
should also be congruent to $3$ mod $4$. The fields
$\mathbb{Q}(\o_p)$ and $\mathbb{Q}(\o_3)$ are linearly disjoint
over $\mathbb{Q}$. Let $\mathbb{K}$ be the unique subfield of
$\mathbb{Q}(\o_3)(\o_p)$ of degree $n_1$ over $\mathbb{Q}(\o_3)$
and let $\mathbb{L} = \mathbb{K}\cdot \mathbb{Q}(\o_{2^{s+1}})$.
Then $\mathbb{L}$ is a cyclic extension of $\mathbb{Q}(\o_3)$, and
the element
$$\gamma = \frac{\pi_1}{\pi_1^{*}}$$ is an (algebraic)
unit-magnitude element that is a non-norm element for the
extension $\mathbb{L}/\mathbb{Q}(\o_3)$ and is independent of SNR.
(When $n_1=1$, so $s=1$, we take $\mathbb{L} =
\mathbb{Q}(\o_3)(\imath)$, and $\gamma =
\displaystyle{\frac{3+\o_3}{3 + \o_3^2}}$.) The $n\times n$
matrices arising from equations (\ref{eq:distributed_CDA}),
(\ref{eq:G_Matrix}) and (\ref{eq:GammaMatrix}) with the above
choice of $\mathbb{F},\mathbb{L}$ and $\gamma$, and with any
choice of scaled integral basis $\{\beta_i\}_{i=0}^{n-1}$ hence
yield a full-diversity, full-rate code over HEX with non-vanishing
determinant satisfying the additional equal power sharing
constraint.
\end{prop}
\begin{proof} See Appendix \ref{sec:Appendix PerfectCodes NonNorm}
\end{proof}

Consequently, we have constructed algebraic, unit-magnitude
`non-norm' elements, valid for use in perfect codes, for any $n$.
Some examples are given in {Table \ref{tab:gamma_table_perfect}}.

 \begin{table}[h]
\caption{Non-Norm Elements}
\begin{center}
\begin{tabular}{|c|c|}
  \hline
  No. of Antennas & Non-norm `$\gamma$' \\
  \hline
  2  & $(2+\imath)/(1+2\imath)$ \\
    & $(1+4\imath)/(1-4\imath)$ \\
  \hline
  3  & $(3+\o_3)/(3+\o_3^{*})$ \\
    & $(1+9\o_3)/(9+\o_3)$ \\
  \hline
  4  & $(2+\imath)/(2-\imath)$ \\
    \hline
  5  & $(3+2\imath)/(3-2\imath)$ \\
    \hline
  6  & $(3+7\o_3)/(3+7\o_3^{*})$ \\
    \hline
  7  & $(8+5\imath)/(5+8\imath)$ \\
    \hline
  8  & $(2+\imath)/(1+2\imath)$ \\
    \hline
  9  & $(3+\o_3)/(1+3\o_3)$ \\
    & $(4+9\imath)/(9+4\imath)$ \\
    \hline
  \hline
\end{tabular}
\end{center}
\label{tab:gamma_table_perfect}
\end{table}

\subsection{Good Constellation shaping \label{ssec:Good Constellation shaping}}
For any dimension $n$, we now proceed to describe the construction
of the unitary matrix $G$ that complies with the cyclic Galois
requirements of the division algebra. The construction method will
involve the embedding of the scaled integral basis of a submodule
of $\mathcal{O}_\mathbb{L}$ over the ring of integers
$\mathcal{O}_\mathbb{F}$ of $\mathbb{F}$, where
$\mathbb{F},\mathbb{L}$ are as in Propositions
\ref{prop:nonNormUnitMagnQAM} and \ref{prop:nonNormUnitMagnHEX}.

We proceed to first construct lattices for any odd dimension
$n_1$, then lattices of dimension $2^s, \ s\in \mathbb{Z}^+$, and
then proceed to combine them in order to give the final desired
lattices for any dimension $n$ over $\mathbb{Q}(\imath)$ or any
dimension that is not a multiple of $4$ over $\mathbb{Q}(\o_3)$.
Without any loss of generality, we will be analyzing the QAM case,
corresponding to $\mathbb{F} = \mathbb{Q}(\imath)$.  Unless we
state otherwise, the same results will also hold for $\mathbb{F} =
\mathbb{Q}(\o_3)$.

\subsubsection{Orthogonal Lattices in a Cyclic Galois Extension
over $\mathbb{Q}$ of Odd Degree
\label{sssec:OrthoLatticesOfOddDegree}} Recently, the authors in
\cite{BayerOggierViterbo}, Section V, give a detailed exposition
of a previous result in \cite{Erez} of an explicit construction of
$q$-dimensional orthogonal lattices that belong in a $q$-degree
cyclic Galois extension $\mathbb{K}'$ over $\mathbb{Q}$, with the
restriction that $q$ be an odd \emph{prime} integer. We here show
that the same construction actually gives $n_1$-dimensional
orthogonal lattices in $\mathcal{O}_\mathbb{K}'$, for \emph{any
odd integer $n_1$}. Moreover, the field $\mathbb{K}'(\imath)$ will
be precisely the field $\mathbb{K}$ of Proposition
\ref{prop:nonNormUnitMagnQAM}.

The steps for constructing an $n_1$-dimensional orthogonal lattice
in $\mathcal{O}_\mathbb{K}$ are as follows:
\begin{itemize}
\item pick a guaranteed to exist odd prime $p\equiv 1$ (mod $n_1$)
 \item let $\o =\o_p = e^{ \frac{2 \pi \imath}{p}}$ \item find
a guaranteed to exist primitive element $r$ of
$\mathbb{Z}^{\divideontimes}_p$ \item for $m=\frac{p-1}{2}$,
create $\alpha = \prod_{k=0}^{m-1} (1-\o^{r^k}) $ where ${r^{p-1}
= 1} $ \item Find a guaranteed to exist $\lambda$ such that
$\lambda(r-1) \equiv 1$ (mod $p$) and let $ { z = \o^\lambda
\alpha (1-\o) } $ \item For $\sigma(\o) = \o^r$, let $x =
\sum_{k=1}^{\frac{p-1}{n_1}} \sigma^{kn_1}(z)$.
\end{itemize}
The element $x$ is hence in the field $\mathbb{K}'$, the subfield
of $\mathbb{Q}(\o)$ fixed by $\sigma^{n_1}$, of degree $n_1$ over
$\mathbb{Q}$.  It is then the case that the following lattice
generator matrix $G_{n_{1}}$ is unitary and the resulting lattice
(which arises from the canonical embedding of the free
$\mathbb{Z}$ module generated by $x/\sqrt{p}$,
$\sigma(x)/\sqrt{p}$, $\dots$, $\sigma^{n_1-1}(x)/\sqrt{p}$ in
$\mathbb{R}^n_1$) is orthogonal.
\begin{eqnarray} \label{eq:TypeI_Lattice_Generator}
G_{n_1} \! = \! \frac{1}{p}\  \begin{array}{|ccccc|}
x               &  \sigma(x)        & \! \cdots \!         &       \sigma^{n_1-2}(x) &   \sigma^{n_1-1}(x) \\
\sigma(x)   &  \sigma^2(x)  & \! \cdots    \!          &   \sigma^{n_1-1}(x) & x                             \\
\sigma^2(x) & \sigma^3(x)   &   \! \cdots   \!         &   x                               & \sigma(x)             \\
& \vdots        &                                                         &   \vdots                  &                                   \\
\sigma^{n_1-1}(x) & x                 &  \!  \cdots  \! &
\sigma^{n_1-3}(x) & \sigma^{n_1-2}(x)
\end{array}
\end{eqnarray}
 Since $\mathbb{Q}(\o)$ and $\mathbb{Q}(\imath)$ are linearly
disjoint over $\mathbb{Q}$, the field $\mathbb{K} =
\mathbb{K}'(\imath)$ will be cyclic over $\mathbb{Q}(\imath)$, and
the elements $x/\sqrt{p}$, $\sigma(x)/\sqrt{p}$, $\dots$,
$\sigma^{n_1-1}(x)/\sqrt{p}$ will be a scaled integral basis for
$\mathbb{K}/\mathbb{Q}(\imath)$.\\
\begin{proof}
See Appendix \ref{sec:appendix_Proof_Orthogonality_Lattice_Odd}.
\end{proof}

More specifically the first row of $G_{n_1}$ is given by
\[G_{n_1}(0,j) = \frac{1}{p} \o^\lambda \alpha \sum_{k=1}^\frac{p-1}{n_1}
(-1)^{kN+j} (1-\o^{r^{kN+j}}) , \ j=0,..,n_1-1\] and the rest of
the circulant matrix by:
\[G_{n_1}(i+1,j) = G_{n_1}(i,j+1 \mod n_1), \ \ i=0,\cdots,n_1-2.\]
\begin{ex} The first row of the $9$-dimensional $G_9$ is
$\frac{1}{19}\Bigl(
   -2.831 \  7.298 \  -1.435 \ 4.149 \ -8.688 \  -8.451 \   -6.414 \  5.355
   \ -7.983   \Bigr)$ and every next row is obtained by a single left cyclic shift of the previous
   row.  The matrix was obtained by setting $n_1=9, \ p=19, \ r=3$ and $\lambda =
   10$.  Similarly the first row of the $15$-dimensional $G_{15}$
   is $\frac{1}{31}\Bigl(   -2.242 \ 6.361 \  -10.78 \    -8.071 \    7.253 \
   -9.45 \
    1.127 \
   -3.334 \
    8.806 \
   -4.391 \
   10.442 \
    5.404 \
  -11.12 \
  -11.004 \
   -9.989 \
\Bigr)$ obtained by setting $n_1=15, \ p=31, \ r=3$ and $\lambda =
16$.\end{ex}

\subsubsection{Lattices of dimension $m = 2^s$ \cite{BoutrosViterbo98}} For when
the information set is QAM, then $\mathbb{F} = \mathbb{Q}(\imath)$
and we consider $\mathbb{K} = \mathbb{Q}(\o_M)$ where $M =
2^{s+2}$ and $\o_M = \o = e^{2 \pi \imath/M}$ the $M^{th}$
primitive root of unity.  $\mathbb{Q}(\o)$ is a cyclic Galois
extension over $\mathbb{Q}(\imath)$. Considering that the order of
$5$ in
$\mathbb{Z}^{\divideontimes}_{M}\tilde{=}Gal(\mathbb{K}/\mathbb{Q})$
is $m = 2^{s} = \frac{\phi(M)}{2}$, we see that for $\sigma \in
Gal(\mathbb{Q}(\o)/\mathbb{Q})$ such that $\sigma(\o) = \o^5$, it
is the case that $\sigma(\imath) = \sigma(\o^{2^s}) = \o^{2^s5} =
\o^{(1+4)2^s} = \o^{2^s}\o^{2^{s+2}} = \o^{2^s} = \imath$ which
gives that $Gal(\mathbb{K}/\mathbb{Q}(\imath)) = <\sigma>$. Taking
$\{\o^0,\o^1,\o^2,\cdots,\o^{m-1}\}$ to be the integral basis over
$\mathbb{Q}(\imath)$, the canonical embedding then gives the
lattice generator matrix
\begin{equation} \label{eq:canonicalEmbeddingCyclotomic}
G_e = \frac{1}{\sqrt{m}} \biggl[ \sigma^k(\o^i)\biggr]_{i,k} =
\frac{1}{\sqrt{m}} \biggl[ \o^{i\cdot5^k}\biggr]_{i,k}
\end{equation}
The fact that the lattice corresponds to the ring of integers of
the $m$-dimensional cyclic Galois extension $\mathbb{K}$ over
$\mathbb{Q}(\imath)$, allows for $G_e$ to be directly used in
(\ref{eq:distributed_CDA}) to construct the $m\times m$ space-time
code.

Now for $r_i = [1 \ \o^{5^i} \ \o^{5^i2} \ \o^{5^i3} \ \cdots \
\o^{5^i(n-1)}  ]$, $i=0,1,\cdots,m-1$, being the $i^{th}$ row of
$\sqrt{m}G_e^T$ in (\ref{eq:canonicalEmbeddingCyclotomic}), we
have that $r_ir_j^\dag = \sum_{k=0}^{m-1} \o^{5^ik} \o^{5^jk} =
\sum_{k=0}^{m-1} \o^{k(5^i-5^j)}$.  Since $5$ has order
$\frac{M}{4} = \frac{\phi(M)}{2}$ in
$\mathbb{Z}^{\divideontimes}_M$, then $5^i \neq 5^j \ \forall
i\neq j, \ i,j =0,1,\cdots,\frac{M}{4}-1$. This combines with the
fact that $k(5^i-5^j) = k5^j(5^{i-j}-1)\equiv 0$
 (mod $4$) so that each summand pairs with another summand in the
summation so that their ratio is $\o^4$. This symmetry, the fact
that $\frac{M}{2}\equiv 0$ (mod $4$) and the fact that
$\o^{5^i}+(\o^{5^i})^\frac{M}{2} = 0$, means that each summand
$\o^{5^i}$ has another summand as its additive inverse. Together
with the fact that the complex conjugate of $\o$ is $\o^{-1}$,
results in $r_ir_j^\dag = m\delta_{i,j}$ and in the desired
orthogonality $G_eG_e^\dag = I$.  The lattices apply only for
codes over QAM.

\subsubsection{Combining lattices} We will need the
following, which is an easy modification of Proposition 6 in
\cite{BayerOggierViterbo} and which eventually guarantees for the
creation of lattices over a cyclic Galois extension for any
dimension $n$ over $\mathbb{Q}(\imath)$, and any dimension that is
not a multiple of $4$ over $\mathbb{Q}(\o_3)$.
\begin{lem} \label{lem:CompositeLattices}
Let $\mathbb{L}$ be the compositum of $l$ Galois extensions
$\mathbb{K}_i$ over $\mathbb{Q}$ of co-prime degrees $n_i$.
Assuming that there exists an orthogonal
$\mathcal{O}_{\mathbb{K}_i}$-lattice generator matrix $G_i$ for
all $i=1,2,\cdots,l$ then the Kronecker product of these matrices
is a unitary generator matrix of an $n$-dimensional lattice in
$\mathcal{O}_\mathbb{L}$, $n = \prod_{i=1}^l n_i$.
\end{lem}

For this, the discriminants are not required to be coprime since
the involved fields already have coprime degrees, so their
composite is their tensor product over $\mathbb{Q}$. Specifically,
for $\mathbb{F} = \mathbb{Q}(\imath)$, for any $n = n_12^s$, $n_1$
odd, the orthogonal lattice generator matrix $G$ is the Kronecker
product of the generator matrix of the $n_1$-dimensional lattice
from Section \ref{sssec:OrthoLatticesOfOddDegree} and that of the
cyclotomic lattice of dimension $2^s$.  For $\mathbb{F} =
\mathbb{Q}(\o_3)$, for $n = n_1$ odd we again use the
$n_1$-dimensional lattice from Section
\ref{sssec:OrthoLatticesOfOddDegree}, and for $n = 2 n_1$, $n_1$
odd, the orthogonal lattice generator matrix $G$ is the Kronecker
product of the generator matrix of the $n_1$-dimensional lattice
from Section \ref{sssec:OrthoLatticesOfOddDegree} and matrix $C_2
= \begin{array}{|cc|}1 & \imath \\ 1 & -\imath \end{array}$.

For $\mathbb{F} = \mathbb{Q}(\o_3)$, for $n = n_1$ odd we again
use the $n_1$-dimensional lattice from Section
\ref{sssec:OrthoLatticesOfOddDegree}, and for $n = 2 n_1$, $n_1$
odd, the orthogonal lattice generator matrix $G$ is the Kronecker
product of the generator matrix of the $n_1$-dimensional lattice
from Section \ref{sssec:OrthoLatticesOfOddDegree} and the matrix
$C_2 =
\begin{array}{|cc|}1 & \imath \\ 1 & -\imath \end{array}$, coming from the
field $\mathbb{Q}(\imath)$.

The above orthogonal lattice generator matrices correspond to a
suitable scaled integral basis of the $n$-dimensional cyclic
Galois extension $\mathbb{L}/\mathbb{F}$, defined (respectively)
in Propositions \ref{prop:nonNormUnitMagnQAM} and
\ref{prop:nonNormUnitMagnHEX}. As discussed above, these matrices
allow for  good constellation shaping.  Consequently, with this
choice of lattice generator matrix and the choice of
$\mathbb{L}/\mathbb{F}$ and $\gamma$ as in Propositions
\ref{prop:nonNormUnitMagnQAM} and \ref{prop:nonNormUnitMagnHEX},
the code defined by equations (\ref{eq:element of maximal
field}-\ref{eq:GammaMatrix}) form perfect codes satisfying
full-diversity, full-rate, non-vanishing determinant, equal power
sharing, \textit{and} good constellation shaping.

\section{Information theoretic interpretation and generalization of the perfect code conditions}
The D-MG tradeoff \cite{ZheTse} bounds the optimal performance of
a space-time code ${\cal X}$ operating at rate $R$ bpcu,
corresponding to a multiplexing gain
$$r = \frac{R}{\log_2(\text{SNR})}.$$  The {\em diversity gain}
corresponding to a given $r$, is defined by
\[
d(r) \ = \ - \lim_{\text{SNR} \rightarrow \infty }
\frac{\log(P_{e})}{\log (\text{SNR})} ,
\]
where $P_{e}$ denotes the probability of codeword error.  For the
Rayleigh fading channel, Zheng and Tse \cite{ZheTse} described the
optimal tradeoff between these two gains by showing that for a
fixed integer multiplexing gain $r$, the maximum achievable
diversity gain is
\begin{equation}
d(r) \ = \ (n-r)(n_r-r). \label{eq:Zheng-Tse}
\end{equation}
The function for non-integral values is obtained through
straight-line interpolation.

We use this D-MG approach as a basis for interpreting and
generalizing the conditions that define perfect-codes.
\subsubsection{Full rate condition}
Consider an $n\times T$ code $\mathcal{X}$ where each code-matrix
carries $m$ information symbols per channel use from a discrete
constellation $\mathcal{A}$ such as QAM.  It is then the case that
 \[|\mathcal{X}| = 2^{RT} =  2^{rT\log_2\text{SNR}} =
\text{SNR}^{rT} = |\mathcal{A}|^{mT}\] which implies that
$|\mathcal{A}| \dot = \mbox{SNR}^\frac{r}{m}$ and since the
constellation is discrete, we have that $\mathbb{E}[\|\alpha\in
\mathcal{A}\|^2]\dot = |\mathcal{A}|$. The fact that each element
$X_{i,j}$ of a code matrix is a linear combination of elements of
$\mathcal{A}$, gives that $$\mathbb{E}[\| X_{i,j} \|^2] \ \dot = \
|\mathcal{A}| = \mbox{SNR}^\frac{r}{m}.$$  The SNR normalizing
factor $\nu$ that guarantees that $\mathbb{E}[\|\nu H X\|_F^2] =
\mathbb{E}[\|\nu X\|_F^2] \dot = \text{SNR}$ is then given by
\begin{equation}\label{eq:normalization Factor Perfect Code TxT}
\nu^2 \ \dot = \  \mbox{SNR}^{1-\frac{r}{m}}.\end{equation}
Without loss of generality we can assume that there exist two
code-matrices $X_1, X_2\in \mathcal{X}$, with each $X_i$ mapping
the information $nm$-tuple $\{\alpha_i,0,0,\cdots,0\}$, where
$\alpha_i \ \dot = \ \mbox{SNR}^0$.  As a result, the determinant
and trace of the difference matrix $\Delta X$, is a polynomial of
degree less than $n$ over $\alpha = \alpha_1-\alpha_2 \ \dot = \
\text{SNR}^0$, with coefficients independent of SNR, i.e.
$$\det(\Delta X \Delta X^\dag ) \ \dot = \ \text{Tr}(\Delta X \Delta X^\dag)
\ \dot =  \ \mbox{SNR}^0$$ and thus with all its eigenvalues $$l_i
\ \doteq \ \mbox{SNR}^0.$$ The corresponding pairwise error
probability $\text{PEP}(X_1 \rightarrow X_2)$, in the Rayleigh
fading channel \cite{TarSesCal,GueFitBelKuo}, then serves as a
lower bound to the codeword error probability $P_e$, i.e.,
\begin{eqnarray*} P_e &\geq & \mbox{PEP}(X_1 \rightarrow X_2) \ \doteq \
\frac{1}{\prod_{j=1}^n [1 + \frac{\theta^2}{4} l_j ]^{n_r}} \\
& \doteq &\mbox{SNR}^{-n_rn (1- \frac{r}{m})} \end{eqnarray*}
which results in a diversity gain of \begin{eqnarray*}d(r) \ \leq
n_rn \left(1- \frac{r}{m} \right).
\end{eqnarray*}
What this means is that $m$ discrete information symbols per
channel use can potentially sustain reliable communication for up
to rate $R_{\max} \approx m\log_2(\text{SNR})$.

For large SNR, the outage capacity over an $n\times n_r$ Rayleigh
fading channel is given by $C_\text{out} \ \approx \ \min \{n,n_r
\} \log_2(\text{SNR})$, implying a maximum achievable multiplexing
gain of ${r_{\max} = \min \{ n,n_r\}}$.  Consequently the relation
between $R_{\max}$ and $C_\text{out}$, allows for the
interpretation that the full rate defining condition is necessary
for reliable transmission at rates close to the outage capacity of
the Rayleigh fading channel, independent of the channel topology.
Equivalently, given some rate $R$, the full rate defining
condition is necessary for reliable transmission at the smallest
allowable SNR
$$\mbox{SNR}_{\min} \ \dot = \  2^{\frac{R}{\min(n,n_r,m)}}$$ again independent of the channel topology.
Let us now re-examine the full-rate condition, in conjunction with
the determinant condition.
\subsubsection{Non-vanishing determinant condition}
We consider the ${n}\times T$ truncated code $\mathcal{X}$, $T\geq
{n}$, constructed by deleting the same $T-n$ rows from all the
code-matrices $X^{'}$ of a $T\times T$ perfect code
$\mathcal{X}^{'}$.    We have seen that for an $n\times n$ code
mapping $n^2$ information symbols from a discrete constellation
($n$ information symbols per channel use), the standard
$n$-dimensional `folding' ($|\mathcal{X}| = |\mathcal{A}|^{n^2}$)
forces a normalizing factor of $\nu^2 =
\text{SNR}^{1-\frac{r}{n}}$, whereas in the truncated $n\times T$
code mapping $T^2$ information symbols ($\frac{T^2}{n}$
information symbols per channel use), the constellation is folded
in $T$ dimensions ($|\mathcal{X}| = |\mathcal{A}|^{T^2}$),
requiring for
$$\nu^2 = \mbox{SNR}^{1-\frac{r}{T}}.$$
This scenario accentuates the fact that in essence, we are limited
by a lower bound on the determinant of the energy-normalized
difference matrix $\nu^2 \Delta X \Delta X^\dag$.
 As a result, for
the $n$-dimensional case, the defining condition of non-vanishing
determinant for the non-normalized matrix $\Delta X \Delta
X^\dag\geq \mbox{SNR}^0$, translates to
\begin{eqnarray*}\det[\nu^2 \Delta X \Delta X^\dag]& \geq &
(\nu^2)^{n} \mbox{SNR}^0 = (\mbox{SNR}^{1-\frac{r}{n}})^{n}\\& = &
\mbox{SNR}^{n-r}\end{eqnarray*} which, for the $n\times T$ case
with $T$-dimensional folding, translates back to the determinant
bound
$$\det(\Delta X \Delta X^\dag)\geq
\mbox{SNR}^{-\frac{r}{T}(T-n)}$$ for the non-normalized
code-matrices.  But from \cite{EliRajPawKumLu,isit05_explicit}, we
see that the above determinant bound is the best that any code can
attain, thus allowing us to generalize the full-rate,
full-diversity and non-vanishing determinant perfect code
conditions, to the general condition of having
\begin{equation}\label{eq:generalized DMG optimality conditions}
\det[\nu^2 \Delta X \Delta X^\dag]  \ \dot \geq \
\text{SNR}^{n-r}, \ 0\leq r \leq \min(n,n_r). \end{equation}

In regards to non-minimum delay perfect codes, let us briefly note
that codes resulting from row deletion of perfect codes
essentially maintain all the conditions of the original
minimum-delay perfect code constructions except that now the
vectorization of the code-matrices is not isometric to
$\text{QAM}^{T^2}$.   Non-minimum delay perfect codes can be
constructed though for delays $T = nk, \ k\in \mathbb{Z}^+$ that
are multiples of $n$, by the horizontal stacking construction
found in \cite{EliRajPawKumLu,isit05_explicit} which maintains the
non-vanishing determinant property as well as the isometry of the
code matrices with $\text{QAM}^{n^2k}$.

Let us now incorporate all the perfect-code defining conditions in
order to provide an information theoretic interpretation that
spans both the high and the low SNR regimes.
\subsubsection{Approximate universality, information
losslessness and Gaussian-like signalling} We begin with:
\begin{thm} Perfect codes are both approximately universal as well as information lossless.\label{thm:approximate
universality and information losslessness}\end{thm}
\begin{proof}See Appendix \ref{sec:appendix_Proof_ApproxUniversal And Information Lossless Perfect
Codes}.\end{proof} The code's information losslessness, shown in
the proof to be the result of the CDA structure and the last two
conditions, essentially allows for the code to maintain the
maximum mutual information corresponding to the channel and
signalling set statistics.  This mutual information is empirically
related to the Gaussian-like signalling set and its good
covariance properties, observed in Figure \ref{fig:Covariance for
capacity achieving arguments}.
\begin{figure}[h]
\begin{center}
\includegraphics[angle=-90,width=0.32\columnwidth]{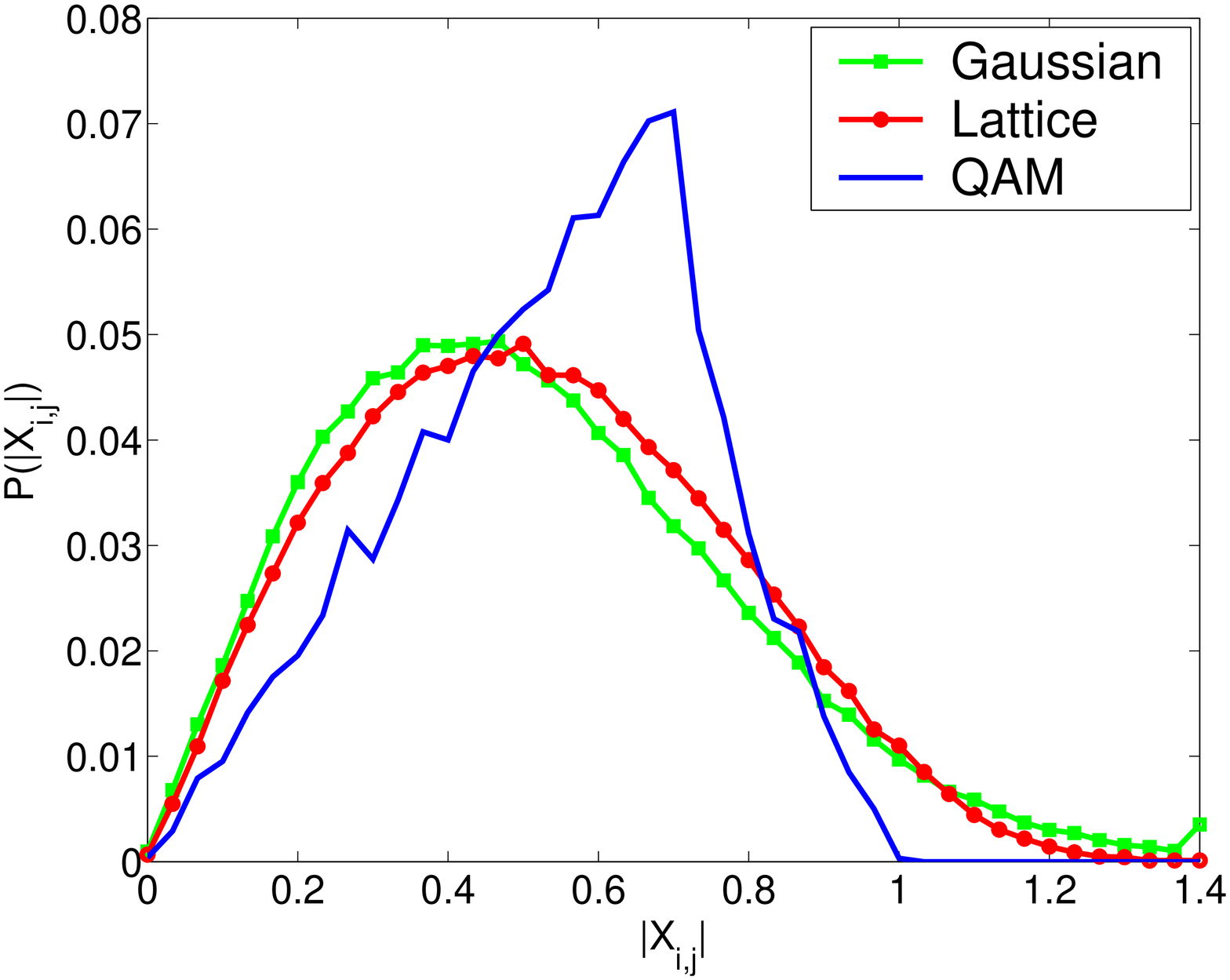}
\includegraphics[angle=-90,width=0.32\columnwidth]{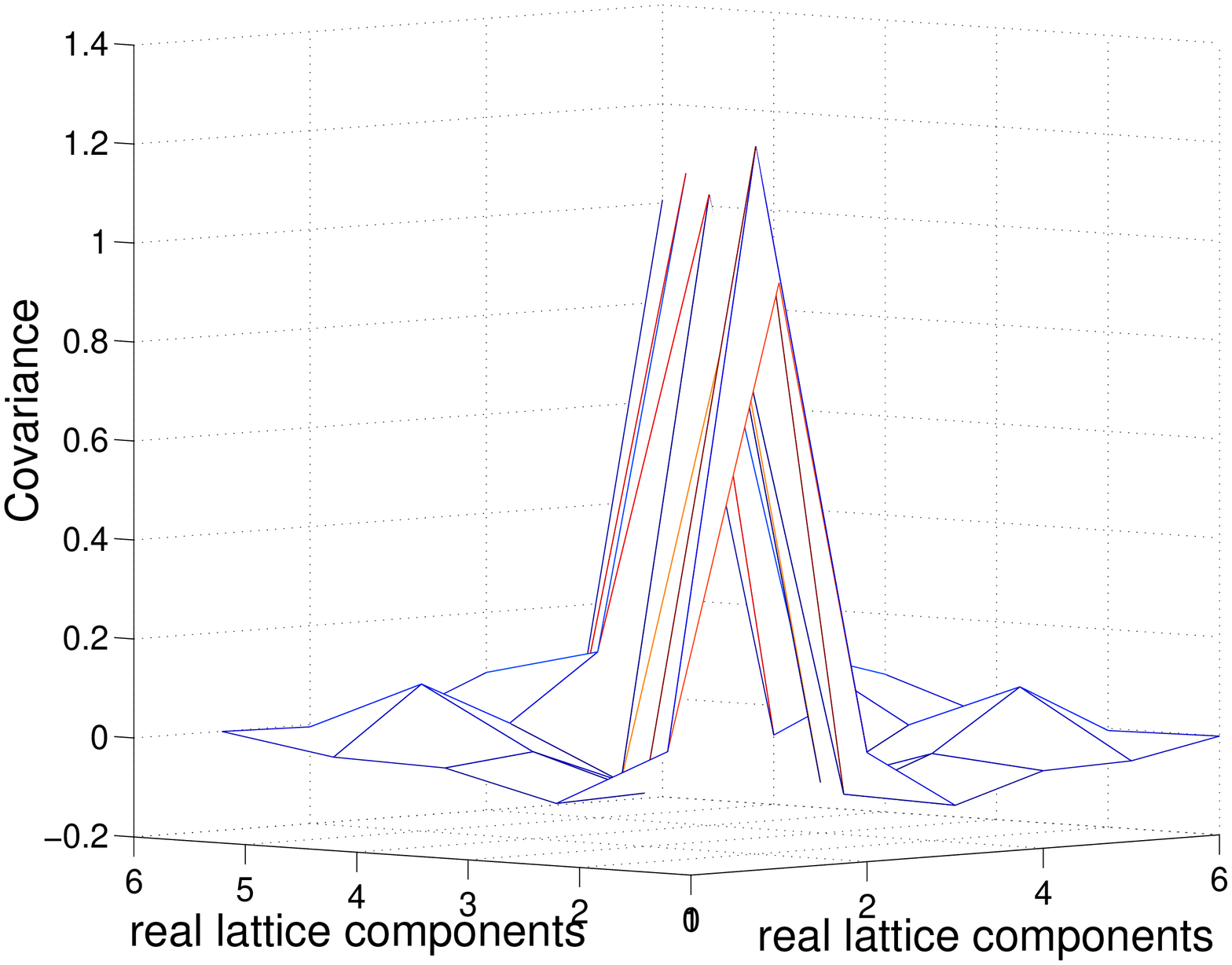}
\includegraphics[angle=-90,width=0.32\columnwidth]{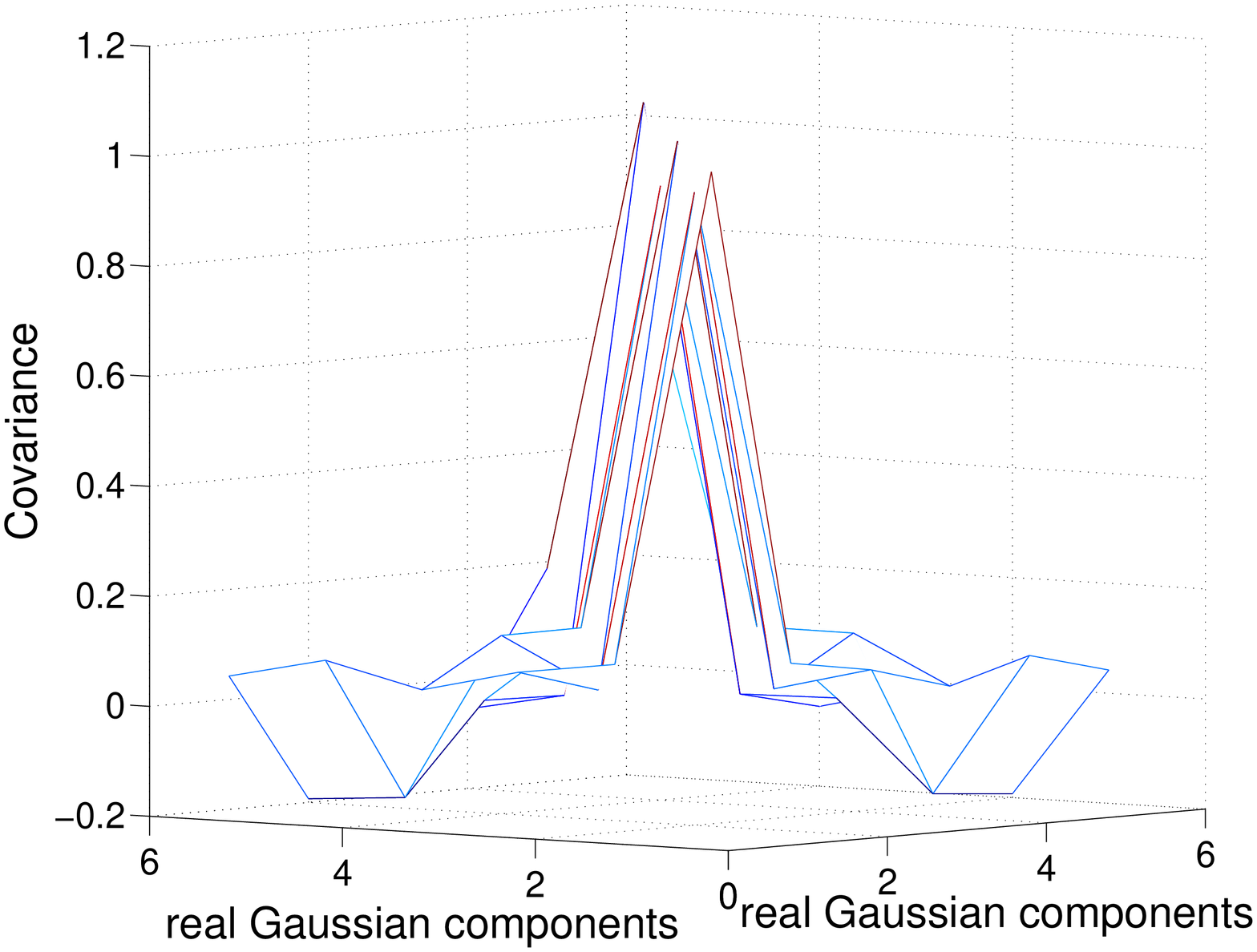}
\caption{Gaussian nature of the signalling set of the $3\times 3$
perfect code, compared to QAM and random Gaussian signalling
(left). Covariance of columns of perfect codes (center) compared
to covariance of random Gaussian vectors (right).
\label{fig:Covariance for capacity achieving arguments}}
\end{center}
\end{figure}
The expedited rate with which the signalling becomes Gaussian,
relates to the high-dimensional and orthogonal nature of the
lattice generator matrix which together with a unit-magnitude
non-norm element, jointly allow equal magnitudes for the diagonal
elements of the covariance matrix of the signalling set.

Let us now draw from the information theoretic interpretation of
the defining conditions and provide variants of perfect codes that
are specifically tailored for channels with a smaller number of
receive antennas, and which manage to maintain good performance at
a considerably reduced sphere decoding complexity.
\subsection{Channel topology and efficient variants of perfect codes}
We have seen that $n\times n$ perfect codes utilize $n$ different
layers to achieve approximate universality for all $n_r$.  Each
layer has non-vanishing product distance and maps $n$-elements
from a discrete constellation, thus maintaining two properties
that were shown in \cite[Theorem 4.1]{TavVisUniversal_2005} to
guarantee for optimality over the statistically symmetric parallel
channel, i.e. a channel with a diagonal fading coefficient matrix,
as well as potentially allowing for optimality over the
statistically symmetric $n\times 1$ MISO channel.  To offer
intuition, we observe that the sum-capacity of the $n_r$
independent MISO channels relates to the full rate condition,
whereas the achieved full diversity relates to the CDA structure
and the discreteness of the powers of $\Gamma$ which manage to
translate the non-vanishing product distance to an overall
non-vanishing determinant, and thus to keep the different layers
independent and at some non-vanishing distance from each other.
The full rate condition comes with a sphere decoding complexity of
$O(n^2)$, but as the number of MISO channels reduces with $n_r$,
so does the required decoding complexity. Codes over such channels
can have the form
\begin{eqnarray*} \label{eq:codematrix with layers of smaller dimension} X &= &\sum_{j=0}^{n-1}
\Gamma^j \biggl( diag \bigl( \underline{f}_{j}\cdot T \cdot G
\bigr) \biggr) \end{eqnarray*} where $T(i,j)  =  1, \ i=j\in
[1,..,n_r]  \ \ \text{and} \ T(i,j)=0 \ \text{otherwise}$, or can
have the form
\begin{eqnarray*} \label{eq:codematrix with fewer layers} X &= &\sum_{j=0}^{n_r-1}
\Gamma^j \biggl( diag \bigl( \underline{f}_{j}\cdot G \bigr)
\biggr). \end{eqnarray*} Note here that the above codes have not
been proven to be D-MG optimal.

\begin{figure}[h]
\begin{center}
\includegraphics[angle=-90,width=0.7\columnwidth]{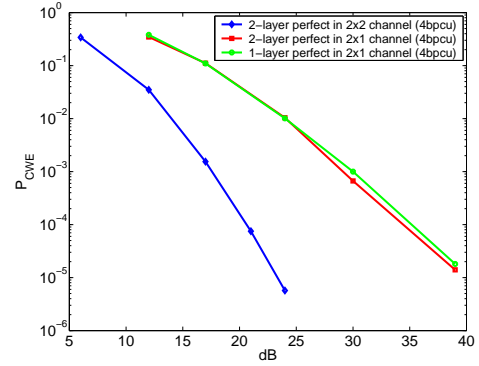}
\caption{For $n=2,n_r = 1$, the single-layer perfect code variant
exhibits similar error performance as the 2-layer (standard)
perfect code.   This changes when a second receive antenna is
added.
\label{fig:data_BayerOurs_in_2x1_vs_IRours_in_2x1_vs_GoldenIn2x2}}
\end{center}
\end{figure}

Motivated by the down-link requirements and by the cooperative
diversity uses of space-time coding in wireless networks, we will
concentrate on the MISO case ($n_r = 1$), for which a D-MG optimal
perfect code variant
\begin{equation}\label{eq:diagonal_restricted_perfect_code}\mathcal{X}_d = \{ diag(\underline{x})
= diag(\underline{f} \cdot G), \ \forall \underline{f}\in
\text{QAM}^n\}.\end{equation} with sphere decoding complexity of
$O(n)$ was recently constructed in
\cite{EliaKumar_DAF_allerton05,EliaKumarRelayJournal} for all $n$,
together with the code \begin{equation}\label{eq:integral
restriction perfect code from perfect code
journal}\mathcal{X}_{ir} = \{ X = \sum_{k=0}^{n-1} f_k \Gamma^k, \
\forall f_k\in \mbox{QAM-HEX} \}
\end{equation}
that corresponds to the center of the division algebra.  With the
exception of $n=2$, $\mathcal{X}_{ir}$ has not yet been proven to
be D-MG optimal. Figure
\ref{fig:data_BayerOurs_in_2x1_vs_IRours_in_2x1_vs_GoldenIn2x2}
provides a performance comparison between the single dimensional
perfect code variant with the equivalent standard perfect-code.

\section{Recent developments involving perfect codes \label{sec:recent developments after perfect codes}}
The proposed high-dimensional perfect codes have had an impact on
establishing outage-based optimality expressions for wireless
networks where independently distributed users utilize different
parts of space-time schemes to relay messages for one another,
hence improving the overall quality of service (\cite{LanemanII}
etc). Up to now, outage-based optimality results were known only
for infinite time duration networks in which the assisting relays
required full knowledge of the channel. Encoding was based on
random Gaussian codes.  Using perfect codes as an information
theoretic tool, it was shown in
\cite{EliaKumar_LPRN_allerton05,EliaKumar_DAF_allerton05} that the
same optimality can be achieved, for finite and minimum delay, and
without requiring knowledge of the channel at the intermediate
relays. This was achieved for the most general network topology
and statistical characterization.

Perfect codes and some perfect-code variants, provided for the
first ever optimal encoding method
\cite{EliaKumar_LPRN_allerton05,Belfiore_Dynamic_AAF_allerton05,EliaKumar_DAF_allerton05,EliaKumarRelayJournal}
in several cooperative-diversity schemes such as the non-dynamic
linear-processing (receive-and-forward) scheme \cite{JingHass04b},
the non-dynamic selection-decode-and-forward scheme
\cite{LanWorTSEIEEE} and finally for the dynamic
receive-and-forward scheme \cite{ElGamalDUplex}.

\section{Examples of new perfect codes and simulations}
\subsection{Examples of new perfect codes}
$ \bullet $ A $2\times 2$ perfect code can be chosen to have
code-matrices which prior to SNR normalization, are of the form
\begin{eqnarray*}X &=& \frac{1}{\sqrt{2}} \,
\begin{array}{|cc|} f_{0,0}+f_{0,1}\o^3_8 &
\gamma(f_{1,0}+f_{1,1}\sigma(\o^3_8))\\
f_{1,0}+f_{1,1}\o^3_8 & f_{0,0}+f_{0,1}\sigma(\o^3_8)
\end{array}\\ \\& = &\frac{1}{\sqrt{2}} \,\begin{array}{|cc|} f_{0,0}+f_{0,1}\o^3_8 &
\gamma (f_{1,0}+f_{1,1}\o^7_8)\\
f_{1,0}+f_{1,1}\o^3_8 & f_{0,0}+f_{0,1}\o^7_8
\end{array}\end{eqnarray*}
where $f_{i,j}$ are from the desired QAM constellation,
$\o_8:=e^{\frac{2\pi \imath}{8}}$ and $\gamma =
\frac{2+\imath}{1+2\imath}$. Matrices map $n^2 = 4$ information
elements from QAM.  Furthermore the signalling set, in the form of
the layer-by-layer vectorization of the code-matrices, before SNR
normalization, comes from the lattice
\begin{multline*}\Lambda =
\bigl\{ [f_{0,0} \ f_{0,1} \ f_{1,0} \ f_{1,1}]R_v \ : \ \\
\forall [f_{0,0},f_{0,1},f_{1,0},f_{1,1}]\in \mbox{QAM}^{n^2}
\bigr\}\end{multline*} where
$$R_v = \frac{1}{\sqrt{2}}\begin{array}{|cccc|}
1 & 1 & 0 & 0 \\
\o_8^3 & \o_8^7 & 0 & 0 \\
0 & 0 & 1 & \gamma \\
0 & 0 & \o_8^3 & \gamma \o_8^7 \end{array}$$ satisfying the
defining condition of
$$R_vR_v^\dag = I_4.$$
 We find the smallest possible determinant,
prior to SNR normalization, to be
\[ \det( \Delta X \Delta X^\dag
)_{\min} = \frac{1}{20}\] which is larger than some previously
constructed $2\times 2$ perfect codes.  The code's performance
improves if the existing $G = \begin{array}{|cc|}1 & 1\\ \o_8^3 &
\o_8^7
\end{array}$ is substituted with
${G_2 =
\begin{array}{|cc|} 0.5257 & 0.8507\\ 0.8507 & -0.5257
\end{array}}$ taken from \cite{BayerOggierViterbo}.\\
Other examples: \bit \item The $5\times 5$ perfect space-time code
is given by \eit
\begin{equation*} \label{eq:5x5PerfectSTcode}
\mathcal{X} = \biggl\{ X = \sum_{j=0}^{4} \Gamma^j \bigl( diag
\bigl( \underline{f}_{j}\cdot G_5 \bigr) \bigr)  , \ \
\underline{f}_{j}\in \mbox{QAM}^5 \biggr\}\end{equation*} for
$\Gamma$ given in (\ref{eq:GammaMatrix}) based on $\gamma =
\frac{3+2\imath}{2+3\imath}$, and generator matrix {\scriptsize
\begin{eqnarray*} \label{eq:idealGenMatr5}
G_5 = \begin{array}{|ccccc|}
  -0.3260  &  0.5485  & -0.4557  & -0.5969  & -0.1699\\
    0.5485 &  -0.4557 &  -0.5969 &  -0.1699 &  -0.3260\\
   -0.4557 &  -0.5969 &  -0.1699 &  -0.3260 &   0.5485\\
   -0.5969 &  -0.1699 &  -0.3260 &   0.5485 &  -0.4557\\
   -0.1699 &  -0.3260 &   0.5485 &  -0.4557 &  -0.5969\\
\end{array}
\end{eqnarray*}}
\normalsize \bit \item The $7\times 7$ perfect space-time code is
given by\eit
\begin{equation*} \label{eq:7x7PerfectSTcode}
\mathcal{X} = \biggl\{ X  = \sum_{j=0}^{6}  \Gamma^j \bigl( diag
\bigl( \underline{f}_{j}\cdot G_7 \bigr) \bigr) , \ \
\underline{f}_{j}\in \mbox{QAM}^7 \biggr\}\end{equation*} for
$\Gamma$ based on $\gamma = \frac{8+5\imath}{8-5\imath}$, and
generator matrix $G_7=$ {\scriptsize
\begin{eqnarray*} \label{eq:idealGenMatr7}
\begin{array}{|ccccccc|}
   -0.681 &   0.163  & -0.449  &  0.077 &   0.082  &  0.276  & -0.469\\
    0.163 &  -0.449  &  0.077  &  0.082 &   0.276  & -0.469  & -0.681\\
   -0.449 &   0.077  &  0.082  &  0.276 &  -0.469  & -0.681  &  0.163\\
    0.077 &   0.082  &  0.276  & -0.469 &  -0.681  &  0.163  & -0.449\\
    0.082 &   0.276  & -0.469  & -0.681 &   0.163  & -0.449  &  0.077\\
    0.276 &  -0.469  & -0.681  &  0.163 &  -0.449  &  0.077  &  0.082\\
   -0.469 &  -0.681  &  0.163  & -0.449 &   0.077  &  0.082  &  0.276\\
\end{array}
\end{eqnarray*}}
\normalsize \bit \item The $25\times 25$ integral restriction code
\eit is given by\[ \mathcal{X}_{\text{ir}} = \biggl\{ X =
\sum_{k=0}^{24} s_k \Gamma^k , \ \ s_k \in \mbox{QAM} \biggr\} \]
with $\gamma = \frac{3+2\imath}{2+3\imath}$.  This code has the
same sphere decoding complexity of $O(25)$ as the $5\times 5$
standard perfect code in the example above, and is expected to
have the same performance, when $n_r = 1$, as the $25\times 25$
perfect code whose sphere decoding complexity is $O(625)$.

\normalsize

\subsection{Simulations}
\begin{figure}[h]
\begin{center}\includegraphics[angle = -90,width=0.7\columnwidth]{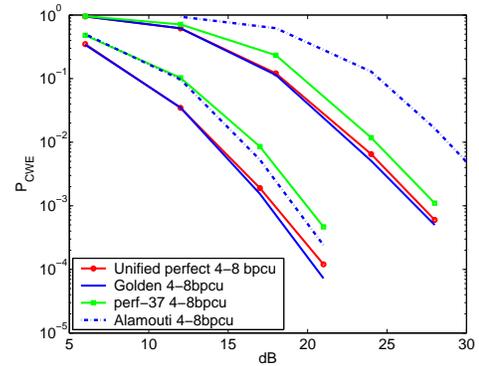}
\caption{Low rate comparison of the unified perfect code with two
perfect codes from \cite{PerfectCodes} and with the Alamouti
code\label{fig:OurBayer vs Golden vs p37 vs Alamouti low SNR}}
\end{center}
\end{figure}

\begin{figure}[h]
\begin{center}\includegraphics[angle = -90,width=0.7\columnwidth]{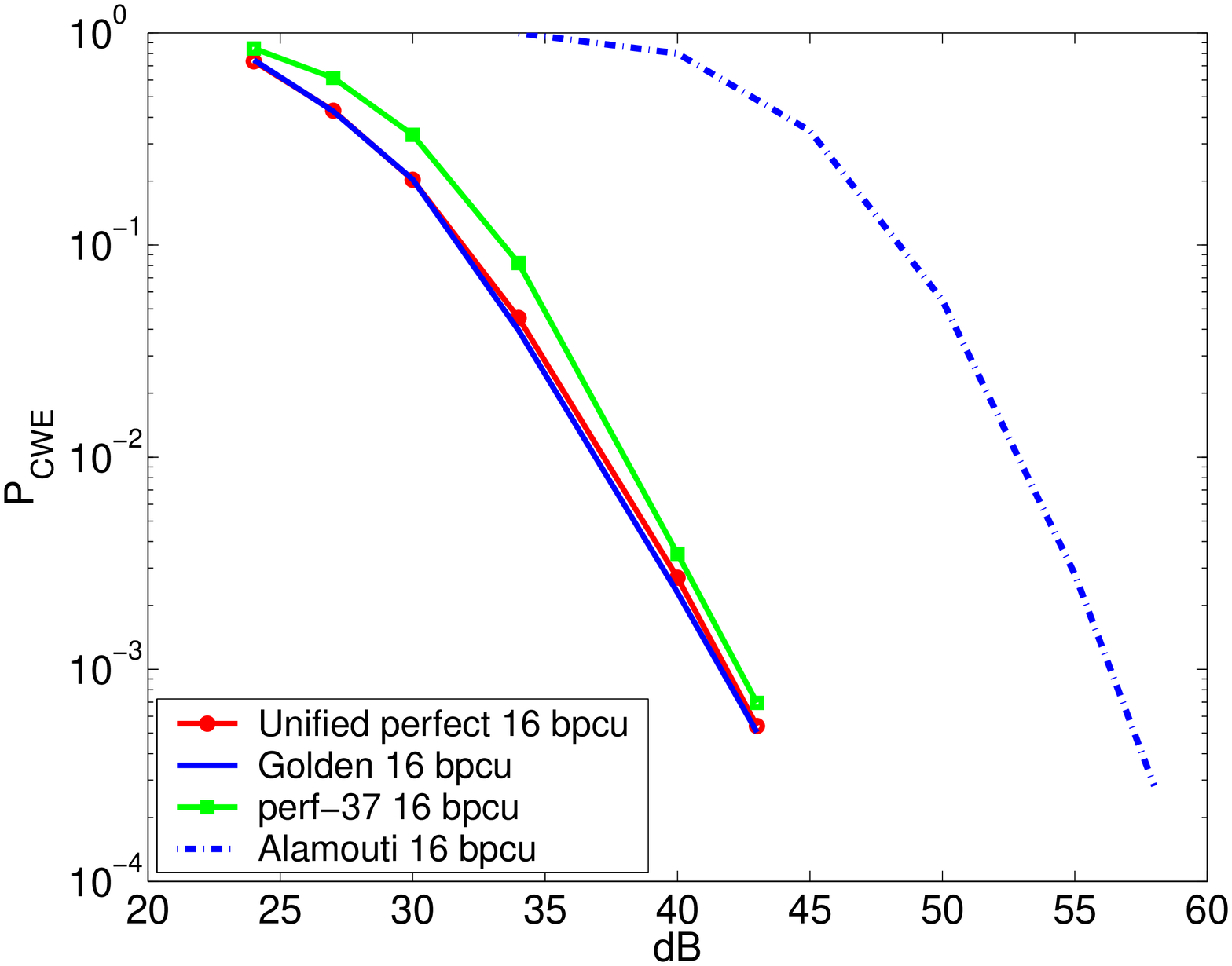}
\caption{High rate comparison of the unified perfect code with two
perfect codes from \cite{PerfectCodes} and with the Alamouti
code\label{fig:OurBayer vs Golden vs p37 vs Alamouti high SNR}}
\end{center}
\end{figure}

\begin{figure}[h]
\begin{center}\includegraphics[angle = -90,width=0.7\columnwidth]{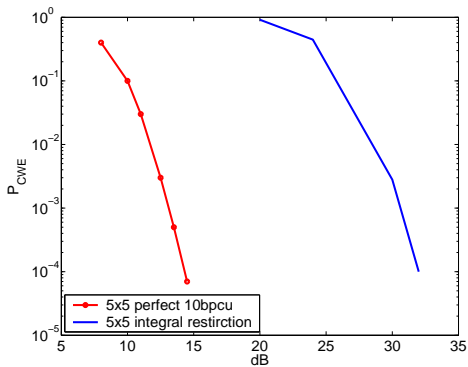}
\caption{Comparison of the $5\times 5$ perfect code with the
$5\times 5$ single dimensional integral restriction code
($n_r=5$).\label{fig:5x5perfect_vs_5x5IRperfect_in5x5Rayleigh_10bpcu}}
\end{center}
\end{figure}

All the simulations assume $\mathbb{C}\mathcal{N}(0,1)$ fading and
thermal noise.  A sphere decoder was used. We begin with
Figure~\ref{fig:Gradual improvement in 3x3 performance} to
indicate the performance improvement as the different defining
conditions are satisfied one-by-one.  The first curve from the top
corresponds to satisfying the full-diversity condition
(commutative CDA code - orthogonal design). The second curve now
includes the full-rate condition (random, full-rate,
linear-dispersion codes). The third curve corresponds to the
family of D-MG optimal but not information lossless CDA codes
presented in \cite{EliRajPawKumLu}, which achieve the first three
criteria of full-diversity, full-rate, and non-vanishing
determinant.  The performance transition from the CDA codes to
perfect codes is described by the next two curves.
\begin{figure}[h]
\begin{center}
\includegraphics[angle=-90,width=0.7\columnwidth]{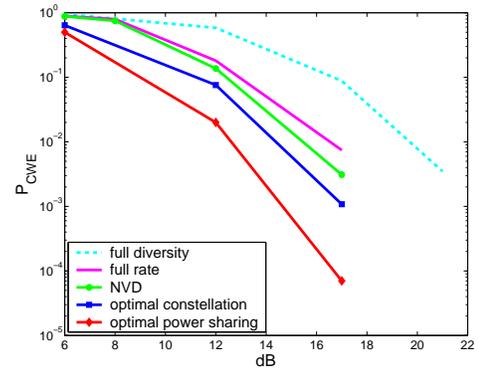}
\caption{Performance improvements attributed to achieving the
different criteria for the perfect codes \label{fig:Gradual
improvement in 3x3 performance}}
\end{center}
\end{figure}
Figures \ref{fig:OurBayer vs Golden vs p37 vs Alamouti low SNR}
and \ref{fig:OurBayer vs Golden vs p37 vs Alamouti high SNR} show
a comparison of the $2\times 2$ unified perfect code presented
here, with some perfect codes from \cite{PerfectCodes} and with
the Alamouti code ($n_r = 2$). The Golden code \cite{BelRekVit}
performs best among all existing $2\times 2$ perfect codes.  When
rates are lower, the unified perfect and the Golden code perform
better than the orthogonal design whereas one of the perfect codes
does not always do so. For higher rates, all considered perfect
codes perform substantially better than the orthogonal design.  At
all rates and all SNR, the perfect code constructed here has
performance very close to that of the Golden code.  In Figure
\ref{fig:5x5perfect_vs_5x5IRperfect_in5x5Rayleigh_10bpcu} we show
the performance of the newly constructed $5$-dimensional perfect
code and compare that with the corresponding $5\times 5$
single-dimensional commutative perfect code (\ref{eq:integral
restriction perfect code from perfect code journal}).  As
expected, the former utilizes fully the $n_r = n=5$ channel and is
thus able to transmit with a small probability of error at high
rates and low SNR.

\section{Conclusion}
\normalsize We have explicitly constructed perfect space-time
codes for any number $n$ of transmit antennas, any number $n_r$ of
receive antennas and any delay $T$ that is a multiple of $n$.
 Achieving all the defining conditions from \cite{PerfectCodes},
allows for perfect codes to exhibit performance that is currently
unmatched. The information theoretic interpretation of the
exhibited good performance both for low and high SNR, is that the
defining conditions jointly endow the code with approximate
universality and the ability to provide for near optimal mutual
information.

High dimensional perfect codes cover a much needed requirement for
optimal codes in multi-user cooperative diversity wireless
networks, where each user acts as a transmit antenna.
Specifically, perfect codes have already been used to establish
the high-SNR outage region of unknown channels, and have provided
the first ever optimal schemes for a plethora of cooperative
diversity methods.

\appendices

\section{Proof of construction methodology for non-norm elements \label{sec:Appendix PerfectCodes NonNorm}}
We will prove here Propositions
\ref{prop:nonNormUnitMagnQAM}-\ref{prop:nonNormUnitMagnHEX}.

For future reference, we first recall three results that relate to
identifying a ``non-norm'' element $\gamma$, i.e, an element
$\gamma \in \mathbb{F}^\divideontimes$ satisfying $\gamma^i \notin
N_{\mathbb{L}/\mathbb{F}}(\mathbb{L}), \ 0<i<n$ for some
$n$-dimensional field extension $ \mathbb{L} $ of $\mathbb{F}$.
\begin{lem}\label{thm:mainBSR} \cite{KirRaj}
Let $\mathbb{L}$ be a degree $n$ Galois extension of a number
field $\mathbb{F}$ and let ${\frak p}$ be a prime ideal in the
ring ${\mathcal O}_\mathbb{F}$ below the prime ideal ${\frak P}
\subset \mathcal{O}_\mathbb{L} $ with norm given by $\| {\frak P}
\| = \| {\frak p} \|^f$, where $f$ is the inertial degree of
${\frak P}$ over $\frak p$. If $\gamma$ is any element of ${\frak
p} \setminus {\frak p}^2$, then $\gamma^i \notin
N_{\mathbb{L}/\mathbb{F}}(\mathbb{L})$ for any $i=1,2, \cdots,
f-1$.
\end{lem}

In order to find a ``non-norm'' element $\gamma$ in $\mathbb{F} =
\mathbb{Q}(\imath)$ ($\mathbb{F} = \mathbb{Q}(\o_3)$), it is
sufficient to find a prime ideal in $\mathbb{Z}[\imath]$
($\mathbb{Z}[\o_3]$) whose inertial degree $f$ in
$\mathbb{L}/\mathbb{F}$ is $f = [\mathbb{L}:\mathbb{F}] = n$. Such
an ideal is said to be inert in $\mathbb{L}/\mathbb{F}$.
\begin{lem}\label{thm:Lemma11} \cite{Rib}
Let $p$ be any odd prime.  Then for any $k\in \mathbb{Z}$,
$\mathbb{Z}^\divideontimes_{p^k}$ is cyclic of order $\phi(p^k)$.
For any integer $f$ dividing $\phi(p^k)$ there exists an $a\in
\mathbb{Z}_{p^k}^{\divideontimes}$ such that $a$ has order $f$ in
$\mathbb{Z}^\divideontimes_{p^k}$.
\end{lem}
\begin{thm} (Dirichlet's theorem) Let $a,m$ be integers such that $1 \leq a \leq
m, \gcd (a,m) = 1$. Then the progression ${ \{ a, a+m, a+2m,
\hdots, a+km, \hdots \} }$ contains infinitely many primes.
\end{thm}

We will now proceed to establish the exact methodology that will
give unit-magnitude non-norm elements $\gamma$, for the different
cases of interest.

\paragraph{Unit-magnitude, non-norm elements for $\mathbb{F} =
\mathbb{Q}(\imath)$} Let \bean n & = & 2^{s} \prod_{i=1}^r
p_i^{e_i} \ = \ 2^{s} n_1 \text{   (say)  } \eean where $p_i$ are
distinct odd primes.  Assume first that $n_1>1$.  Let $p$ be the
smallest odd prime $p$ such that $n_1 \mid (p-1)$. The cyclic
group $\mathbb{Z}_p^{\divideontimes}$ contains an element whose
order equals $(p-1)$. Let $a$ denote this element. Our first goal
is to find a prime $q$ such that
\bean q & = & 5 \pmod{2^{s+2}} \\
 q & = & a \pmod{p}.
\eean Note that \bean q & = & 1 \pmod{4}  \ .\eean Since
$(2^{s+2},p)=1$, we can, by the Chinese Remainder Theorem, find
an integer $b$ such that \bean b & = & 5 \pmod{2^{s+2}} \\
b & = & a \pmod{p}. \eean Note that such an integer $b$ is
relatively prime to $2^{s+2}p$. Consider the arithmetic
progression
\[
b+l(2^{s+2}p), \ \ l=0,1,2,\hdots
\]
By Dirichlet's theorem, this arithmetic progression is guaranteed
to contain a prime $q$ having the desired properties. Now let us
verify that this leads to a CDA.

Let $\mathbb{K}^{'}$ be the  subfield of $\mathbb{Q}(\o_{p})$ that
is a cyclic extension of $\mathbb{Q}$ of degree $n_1$.   Let
$\mathbb{K}$ be the compositum of $\mathbb{K}^{'}$ and
$\mathbb{Q}(\imath)$ and let $\mathbb{L}$ be the compositum of the
fields $\mathbb{K}$ and $\mathbb{Q}(\o_{2^{s+2}})$. Note that
$\mathbb{L}$ is cyclic over $\mathbb{Q}(\imath)$, since it is a
composite of the cyclic extension
$\mathbb{Q}(\o_{2^{s+2}})/\mathbb{Q}(\imath)$ of degree $2^s$ and
the cyclic extension  $\mathbb{K}/\mathbb{Q}(\imath)$ of $n_1$
(note that $2^s$ and $n_1$ are relatively prime). Now consider the
decomposition of the prime ideal $(q)$ in the extension
$\mathbb{L}/\mathbb{Q}$.

Since $q=1 \pmod{4}$ we have that $q$ splits completely in
$\mathbb{Q}(\imath)/\mathbb{Q}$.  Since $q$ has order $(p-1)$ in
$\mathbb{Z}_p$ it follows that $q$ remains inert in
$\mathbb{Q}(\o_p)/\mathbb{Q}$. Since $q=5 \pmod{2^{s+2}}$ and $5$
has order $2^s$ in $\mathbb{Z}_{2^{s+2}}$, it follows that in the
extension $\mathbb{Q}(\o_{2^{s+2}}) / \mathbb{Q}$, $q$ splits
completely in $\mathbb{Q}(\imath)/\mathbb{Q}$ but remains inert
thereafter.

Let $q$ split in $\mathbb{Q}(\imath)/\mathbb{Q}$ according to
\[ q \ = \ \pi_1 \pi_1^*
\]
where $\pi_1=(a+\imath b)$ and $\pi_1^*=(a-\imath b)$. Now by
using the fact that in a field tower
$[\mathbb{E}:\mathbb{K}:\mathbb{F}]$ of field extensions,
$f_{\mathbb{E}/\mathbb{F}}=f_{\mathbb{E}/\mathbb{K}}f_{\mathbb{K}/\mathbb{F}}$,
$g_{\mathbb{E}/\mathbb{F}}=g_{\mathbb{E}/\mathbb{K}}g_{\mathbb{K}/\mathbb{F}}$,
$[\mathbb{E}:\mathbb{F}]=f_{\mathbb{E}/\mathbb{F}} \
g_{\mathbb{E}/\mathbb{F}}$, it follows that $\pi_1$ remains inert
in the extension $\mathbb{L}/\mathbb{Q}(\imath)$.

\begin{figure}[h]
\begin{center}
\includegraphics[width=0.5\columnwidth,height = 0.42\columnwidth]{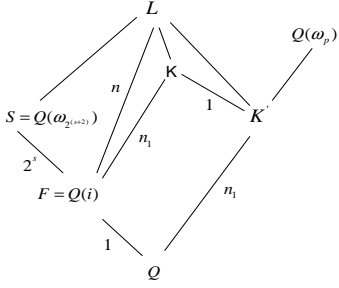}
\caption{The inertial degrees for the maximal field of the CDA
\label{fig:tower_perfect}}
\end{center}
\end{figure}
To now find a non-norm element of unit magnitude, we note that
since the units of $\mathbb{Z}[\imath]$ belong to the set $\{\pm 1
, \ \pm \imath \}$, the associates of \bean \pi_1 & = & a+\imath b
\text{ \ \  belong to the set} \eean
\[
\{a+\imath b, \  \ -a-\imath b, \ \  \imath(a+\imath b), \ \
-\imath(a+\imath b) \}.
\] It follows that since $ab \neq0$, $a-\imath b$ does
not belong to the set of associates of $a+\imath b$. Our goal now
is to show that \[ \gamma \ = \ \frac{\pi_1}{\pi_1^*}
\]
is a non-norm element, i.e.,  that the smallest exponent $k$ for
which $\gamma^k$ is the norm of an element in $\mathbb{L}$, is
$n$. This is the case since if
\[
\gamma^k \ = \ N_{\mathbb{L}/\mathbb{F}}(\ell) \ \ \ \text{some} \
\ell\in\mathbb{L}
\]
then\[ \pi_1^k \ = \ \pi_1^{*k} \prod_{l=0}^{n-1} \sigma^l(\ell) \
\]
where $\sigma$ is the generator of the cyclic Galois group of
$\mathbb{L}/\mathbb{F}$. For $\ell = \frac{a}{b}, \ a,b \in
\cal{O}_{\mathbb{L}}$, we have, in terms of ideals of
$\cal{O}_{\mathbb{L}}$,  $$(\pi_1)^k \prod_{l=0}^{n-1}
(\sigma^l(b)) \ = \ (\pi_1^*)^k \prod_{l=0}^{n-1} (\sigma^l(a)).$$
Since $\sigma(\pi_1) = \pi_1$ we have that if $(\pi_1)$ divides
$(\sigma^l(x))$ for some $l$ and $x\in \cal{O}_{\mathbb{L}}$, it
must divide $(\sigma^l(x))$, for all $l$. This in turn implies
that the power of $(\pi_1)$ in the prime decomposition of
$(\pi_1)^k \prod_{l=0}^{n-1} (\sigma^l(b)) $ is $k \mod n$ whereas
the power of $(\pi_1)$ in the prime decomposition of $(\pi_1^*)^k
\prod_{l=0}^{n-1} (\sigma^l(a))$ is a multiple of $n$.
Equivalently $k$ must be a multiple of $n$.

When $n_1 = 1$, it is sufficient to take $q$ to equal $5$, and
$\mathbb{L}= \mathbb{Q}(\o_{2^{s+2}})$. The prime $5$ splits in
$\mathbb{Q}(\imath)$ as $(1+2\imath)(1-2\imath)$ and then each of
$(1+2\imath)$ and $ (1-2\imath)$ remain inert in the extension
$\mathbb{L}/\mathbb{Q}(\imath)$.  The element
$$
\gamma = \frac{1+2\imath}{1-2\imath}
$$
is then a non-norm element for this extension, for the same
reasons as above.
 This concludes the proof of Proposition
\ref{prop:nonNormUnitMagnQAM}. \epf

\paragraph{Unit-magnitude, non-norm elements for $\mathbb{F} =
\mathbb{Q}(\o_3)$}  Let $n = 2^{s}n_1, \ s\in\{0,1\}$ where $n_1$
is odd. The proof is similar to when $\mathbb{F} =
\mathbb{Q}(\imath)$. Assume first that $n_1 > 1$. We find a prime
$p\equiv 1 \ (\text{mod} \ n_1)$, $p>3$ and a prime $q \in
\mathbb{Z}$, $q\equiv 1 \ (\text{mod} \ 3)$, with
$\text{ord}(q)|_{\mathbb{Z}_{p^{\divideontimes}}} = n_1$.  If
$s=1$, we also require that $q\equiv 3\ (\text{mod} \ 4)$. Assume
that we have found such a $p$ and $q$. The argument for the rest
of the statements in this paragraph are all exactly as in the case
when $\mathbb{F} = \mathbb{Q}(\imath)$:
 The conditions $\text{ord}(q)|_{\mathbb{Z}_{p^{\divideontimes}}} = n_1$
 and $q\equiv 3\ (\text{mod} \ 4)$ (if $s = 1$) guarantee that the
 prime $q$  remains inert in the ring of
integers $\mathcal{O}_{\mathbb{L}'}$ of the cyclotomic field
$\mathbb{L}' = \mathbb{K}(\o_{2^{s+1}})$, where $\mathbb{K}$ is
the unique subfield of degree $n_1$ in the extension
$\mathbb{Q}(\o_p)/\mathbb{Q}$. The field  $\mathbb{L}'$  is cyclic
over $\mathbb{Q}$ of degree $2^s n_1$.
 Since $q\equiv 1 \ (\text{mod} \ 3)$,
the prime $q$ splits  into two distinct primes $\pi_1,\pi_2$ in
$\mathbb{Z}[\o_3]$ which are conjugates of each other.  Let
$\mathbb{L} = \mathbb{L}'(\o_3)$, which is cyclic over
$\mathbb{Q}(\o_3)$ of degree $2^s n_1$.  Then $\pi_1$ will remain
inert in the extension $\mathbb{L}/\mathbb{Q}(\o_3)$. The element
$\gamma = \frac{\pi_1}{\pi_2} = \frac{\pi_1}{\pi_1^*}$ will then
be a unit-magnitude (algebraic) non-norm element for the extension
$\mathbb{L}/\mathbb{Q}(\o_3)$, and the codes constructed with this
data will then be full-diversity, full-rate, and have
non-vanishing determinant, and of course, will satisfy the equal
power-sharing constraint as $\gamma$ is of unit-magnitude.

What is left is to find $p$ and $q$. The prime $p$ is found using
Dirichlet as in the case where $\mathbb{F} = \mathbb{Q}(\imath)$.
To find $q$, first find an integer $b$ that is simultaneously
congruent to $1$ ($\text{mod}\ 3$), to $m$ ($\text{mod}\ p$),
where $m$ is a generator of $\mathbb{Z}_{p^{\divideontimes}}$, and
(if $s =1$) to $3$ ($\text{mod}\ 4$).  This is possible by the
Chinese Remainder Theorem. Next, find the prime $q$ by applying
Dirichlet's theorem to the arithmetic sequence $b + l(3p)$,
$l=0,1,2,\dots$ if $s=0$ and the sequence $b + l(12p)$,
$l=0,1,2,\dots$ if $s=1$.

When $n_1 = 1$ (so $s=1$), we take $\mathbb{L}$ to be
$\mathbb{Q}(\o_3)(\imath) $, and the prime $q$ to be $7$.  Since
$q$ is congruent to $1$ ($\text{mod}\ 3)$ and to $3$ ($\text{mod}\
4)$, $q$ splits into $3+\o_3$ and $3 + \o_3^2$ in
$\mathbb{Q}(\o_3)$ but remains inert in the extension
$\mathbb{Q}(\imath)/\mathbb{Q}$.  It follows that each of $3+\o_3$
and $3 + \o_3^2$ remain inert in the extension
$\mathbb{L}/\mathbb{Q}(\o_3)$. The element
$$
\gamma = \frac{3+\o_3}{3 + \o_3^2}
$$
will then be a non-norm element for this extension, for the same
reasons as above.
 This concludes the
proof of Proposition \ref{prop:nonNormUnitMagnHEX}. \epf

\section{Orthogonal Lattices in $\mathcal{O}_\mathbb{K}$, Where $\mathbb{K}/\Q$ is Cyclic Galois of Odd Degree \label{sec:appendix_Proof_Orthogonality_Lattice_Odd}}

We here show that the construction in \cite{Erez} (of which a
detailed exposition has been provided in \cite[Section
5]{BayerOggierViterbo}) of lattices that belong in a cyclic Galois
extension $\mathbb{K}$ of \emph{prime} degree $q$ over
$\mathbb{Q}$, actually gives without any modification orthogonal
lattices \emph{for any odd degree $n$}. We will follow the
exposition in \cite{BayerOggierViterbo} closely, retaining even
the notation in \cite{BayerOggierViterbo}, and show that the
proofs there only use the fact that $n$ is odd, and not that it is
an odd prime.

To this end, let $n \ge 3$ be a given odd integer, and fix a prime
$p \equiv 1$ (mod $n$).  Note that the existence of such a $p$ is
guaranteed since the sequence $\{1+dn, \ d=1,2,\cdots \}$, as
shown by Dirichlet, contains infinitely many primes.  Let $\o $ be
a primitive $p$-th root of unity.  Thus, $\Q(\o)$ is cyclic of
degree $p-1$ over $\Q$, and contains the real subfield $\Q(\o +
\oi)$ which is cyclic of degree $(p-1)/2$ over $\Q$. Since $n$
divides $p-1$, there is a unique field $\mathbb{K}$ contained in
$\Q(\o)$ which is cyclic of degree $n$ over $\Q$. This is the
field we will work with. Note that since $n$ is odd, $n$ divides
$(p-1)/2$ as well, so $\mathbb{K}$ is contained in the real
subfield $\Q(\o + \oi)$.

Recall that we are following the notation in
\cite{BayerOggierViterbo}.
 Let $G = Gal(\Q(\o)/\Q)$, with generator
$\sigma$, chosen so that $\sigma(\o) = \o^r$, where in turn, $r$
is a generator of $\mathbb{Z}_p^\divideontimes$.  We let
$m=\frac{p-1}{2}$, and observe that $r^m \equiv -1$ (mod $p$). We
also choose $\lambda$ so that $\lambda(r-1) \equiv 1$ (mod $p$).

We define $\alpha$ by $\alpha = \prod_{k=0}^{m-1} (1 - \o^{r^k})$.
The following result is just a combination of Lemmas 3 and 4 of
\cite{BayerOggierViterbo}, and since they have to do purely with
the cyclotomic extension $\Q(\o)/\Q$ and have nothing to do with
$n$, their proofs remain valid:

\begin{lem} \label{lem:combo} The following equalities hold:
\begin{enumerate}
\item $\sigma(\alpha) = - \o^{p-1} \alpha$ \item
$\sigma(\o^{\lambda} \alpha) - \o^{\lambda} \alpha$ \item
$(\o^{\lambda} \alpha)^2 = (-1)^m p$
\end{enumerate}

\end{lem}

We now define $z = \o^{\lambda} \alpha (1-\o)\in
\mathcal{O}_{\Q(\o)}$, and
$$x = Tr_{\Q(\o)/\mathbb{K}} (z) = \sum_{j=1}^{(p-1)/n} \sigma^{jn}(z).$$
Note that $x$ is in $\mathcal{O}_\mathbb{K}$, as $z$ is in
$\mathcal{O}_{\Q(\o)}$. Observing that $$G_NG_N^T(i,j) =
Tr_{\mathbb{K}/\mathbb{Q}}(\sigma^i(x)\sigma^j(x)),$$ we are
interested in $Tr_{\mathbb{K}/\Q} (x \sigma^t(x))$.  The
following, which is Proposition 2 of \cite{BayerOggierViterbo},
gives us the key to constructing the orthogonal lattice.

\begin{prop} $Tr_{\mathbb{K}/\Q} (x \sigma^t(x)) = p^2 \delta_{0,t}$,
for $t = 0, \dots, n-1$.
\end{prop}

\begin{note} Note that $Tr_{\mathbb{K}/\Q} (\sigma^i(x) \sigma^j(x)) = Tr_{\mathbb{K}/\Q} (x
\sigma^{j-i}(x))$.  Thus, if we embed $\mathcal{O}_\mathbb{K}$ in
$\R^n$ via $a \mapsto v(a)= [a, \sigma(a), \dots,
\sigma^{n-1}(a)]$ (note that $\mathbb{K}$ is a real field), this
Proposition says that the vectors $[v(x), v(\sigma(x), \dots,
v(\sigma^{n-1}(x))]$ are orthogonal to one another.
\end{note}

\begin{proof} For $n$ being odd, we have
\begin{eqnarray*}
  Tr_{\mathbb{K}/\Q} (x \sigma^t(x)) &=& \sum_{a=0}^{n-1} \sigma^a(x\sigma^t(x)) \\
    &=& \sum_ {a=0}^{n-1} \sum_{c,j=1}^{(p-1)/n} \sigma^{a+cn}(z)\sigma^{a+t+jn}(z)
\end{eqnarray*}
and from Lemma \ref{lem:combo}
\begin{multline*}
  Tr_{\mathbb{K}/\Q} (x \sigma^t(x)) = \sum_ {a=0}^{n-1} \sum_{c,j=1}^{(p-1)/n}
  (-1)^{a+cn}
   \o^{\lambda} \alpha(1-\o^{r^{a+cn}})\\ \cdot (-1)^{a+t +jn}
   \o^{\lambda} \alpha(1-\o^{r^{a+t +jn}})
\end{multline*}
We observe that since $n$ is odd, $(-1)^{cn} = (-1)^c$ and
$(-1)^{jn} = (-1)^j$.  Moreover, $(-1)^a (-1)^a = 1$, and $(-1)^t$
is common to the sums above.  By Lemma \ref{lem:combo}, we may
replace $(\o^{\lambda})^2$ by $(-1)^m p$. Thus we find, after
rearranging the sums, that
\begin{multline*}
Tr_{\mathbb{K}/\mathbb{Q}} (x \sigma^t(x)) =  (-1)^t (-1)^m p \sum_{c=1}^{(p-1)/n} (-1)^c \cdot \\
\cdot \biggl[ \sum_{a=0}^{n-1}\sum_{j=1}^{(p-1)/n} (-1)^j (1-\o^{r^{a+cn}}) \\
-  \sum_{a=0}^{n-1} \sum_{j=1}^{(p-1)/n} (-1)^j (\o^{r^{a+t +jn}}
- \o^{r^{a+cn}+r^{a+t +jn}}) \biggr]
\end{multline*}
Now the term $\sum_{j=1}^{(p-1)/n} (-1)^j (1-\o^{r^{a+cn}})$ can
be rewritten as $(1-\o^{r^{a+cn}}) \sum_{j=1}^{(p-1)/n} (-1)^j$.
Since $n$ is odd, $(p-1)/n$ is even, and hence, there are as many
positive as negative terms in the expression $\sum_{j=1}^{(p-1)/n}
(-1)^j$, and thus, the sum becomes zero.  Similarly, the term
$\sum_{c=1}^{(p-1)/n} (-1)^c \sum_{a=0}^{n-1}
(-\sum_{j=1}^{(p-1)/n} (-1)^j (\o^{r^{a+t +jn}})$ becomes zero:
this is because the terms in $\sum_{a=0}^{n-1}
(-\sum_{j=1}^{(p-1)/n} (-1)^j (\o^{r^{a+t +jn}})$ are independent
of $c$, while the term $\sum_{c=1}^{(p-1)/n} (-1)^c = 0$ as
$(p-1)/n$ is even and there as many positive as negative terms. We
thus find
\begin{multline*}
Tr_{\mathbb{K}/\Q} (x \sigma^t(x)) =  (-1)^{t+m} p
\sum_{c=1}^{(p-1)/n} (-1)^c \cdot \\ \cdot \sum_{a=0}^{n-1}
\sum_{j=1}^{(p-1)/n} (-1)^j \o^{r^{a+cn}+r^{a+t +jn}}
\end{multline*}

We now have the following:
\begin{lem} \label{lem:Comb_Identity}
\begin{multline*}
\sum_{c=1}^{(p-1)/n} (-1)^c \sum_{a=0}^{n-1} \sum_{j=1}^{(p-1)/n}
(-1)^j \o^{r^{a+cn}+r^{a+t +jn}}\\=\sum_{d=1}^{(p-1)/n} (-1)^d
\sum_{a=0}^{n-1} \sum_{k=1}^{(p-1)/n} \o^{r^{a+nd +nk}+r^{a+t
+nk}}\\
=  \sum_{d=1}^{(p-1)/n} (-1)^d \sum_{a=0}^{n-1}
\sum_{k=1}^{(p-1)/n} \o^{r^{a+kn}(r^{nd}+r^{t})}
\end{multline*}
\end{lem}
\begin{proof} See Appendix \ref{sec:proof_Of_Comb_Lemma}
\end{proof}
As in \cite{BayerOggierViterbo}, we write
\begin{multline*}
\sum_{d=1}^{(p-1)/n} (-1)^d \sum_{a=0}^{n-1} \sum_{k=1}^{(p-1)/n}
\o^{r^{a+kn}(r^{nd}+r^{t})}\\ = \sum_{d=1}^{(p-1)/n} (-1)^d
  \sum_{s=1}^{(p-1) } \o_{d,t}^s
\end{multline*}
where $\o_{d,t} =\o^{ (r^{nd}+r^{t})}$, and of course,
$$
\sum_{s=1}^{(p-1) } \o_{d,t}^s = \begin{cases}
  p-1 & \text{if $\o_{d,t}=1$,}\\ -1 &\text{otherwise}
\end{cases}
$$

To determine when $\o_{d,t} = 1$, note that this happens (as in
\cite{BayerOggierViterbo}) when $t = nd-m+k_1(p-1)$.  Since $n$ is
odd, $n$ divides $m$, so $n$ must divide $t$.  This forces $t=0$.

We now have $\o_{d,t}=1$ implies $r^{nd}\equiv -1 $ (mod $p$), and
writing $-1$ as $r^m$, yields $nd-m = l (p-1)$ for some $l$. This
then gives $d = (p-1) (2l+1)/2n$, which we may write as $(2l+1)$
times $(p-1)/2n)$ (note again that since $n$ is odd, $n$ divides
$(p-1)/2$).  Since $d$ varies in the range $1, \dots, (p-1)/n$, we
find that $l$ must be zero, that is, $d = (p-1)/2n$. Thus,
$\o_{d,t}=1$ precisely when $t=0$ and $d = (p-1)/2n$.

In particular, when $t\neq 0$ then $\o_{d,t}\neq 1$ and we have
that
\begin{multline*} Tr_{\mathbb{K}/\Q} (x \sigma^t(x)) = (-1)^{t+m} p
\sum_{d=1}^{(p-1)/n} (-1)^d \sum_{s=1}^{(p-1) } \o_{d,t}^s \\=
(-1)^{t+m} p \sum_{d=1}^{(p-1)/n} (-1)^d (-1)\end{multline*}
 Once again, since $n$ is odd,
$(p-1)/n$ is even, so the term $\sum_{d=1}^{(p-1)/n} (-1)^d = 0$.
Thus, for $t\neq 0$, $Tr_{\mathbb{K}/\Q} (x \sigma^t(x)) = 0$.

When $t=0$, we find \begin{multline*}Tr_{\mathbb{K}/\Q}(x
\sigma^t(x))  = (-1)^m p \sum_{d=1, d \neq (p-1)/2n}^{(p-1)/n}
\biggl[ (-1)^d (-1)\\ + (-1)^m p (-1)^{(p-1)/2n} (p-1) \biggr]
\end{multline*}
and the right side then yields $p + p(p-1) = p^2$.  To see this
last fact, consider first the case where $(p-1)/2 $ is even (i.e.,
$p\equiv 1$ mod $4$).  Then, since $n$ is odd, $(p-1)/2n$ is also
even. The sum $\sum_{d=1, d \neq (p-1)/2n}^{(p-1)/n} (-1)^d$
equals $\sum_{d=1}^{(p-1)/n} (-1)^d - (-1)^{(p-1)/2n}$, and we
have already seen that, again because $n$ is odd,
$\sum_{d=1}^{(p-1)/n} (-1)^d = 0$.  Thus the right hand side in
the equation above for $Tr_{\mathbb{K}/\Q}(x \sigma^t(x))$ indeed
yields $p^2$ in this case.  We can similarly deal with the case
when $(p-1)/2 $ is odd (i.e., $p\equiv 3$ mod $4$), to find that
in both cases, indeed $Tr_{\mathbb{K}/\Q}(x \sigma^t(x)) = p^2$
when $t=0$. This proves the Proposition.
\end{proof}

\section{Proof of Lemma \ref{lem:Comb_Identity} \label{sec:proof_Of_Comb_Lemma}}

\def\a{\frac{p-1}{n}}
\def\w{\omega}

We wish to prove:
\begin{multline*}
\sum_{c=1}^\a (-1)^c \sum_{a=0}^{n-1} \sum_{j=1}^{\a} (-1)^j
\w^{r^{a+cn} + r^{a+t+jn}} \\= \sum_{d=1}^\a (-1)^d
\sum_{a=0}^{n-1} \sum_{k=1}^{\a} \w^{(r^{nd}+r^t)r^{a+nk}}
\end{multline*}
Set $m = \frac{p-1}{n}$ and denote $\mathbb{Z}/m\mathbb{Z}$ by
$\mathbb{Z}_m$.    In the above equation, the dependence on
$c,j,d,k$ is only through their values $(\text{mod} \ m)$ or
through their values  $(\text{mod} \ 2)$.  If we assume $2|m$,
which follows from the assumption that $n$ is odd, we can then
treat $c,j,d,k$ as elements of $\mathbb{Z}_m$.  We thus have
\begin{multline*}
\sum_{c=1}^\a (-1)^c \sum_{a=0}^{n-1} \sum_{j=1}^{\a} (-1)^j
\w^{r^{a+cn} + r^{a+t+jn}} \\= \sum_{c\in \mathbb{Z}_m} (-1)^c
\sum_{a=0}^{n-1} \sum_{k\in \mathbb{Z}_m} (-1)^k \w^{r^{a+cn} +
r^{a+t+kn}}
\\=\sum_{c\in \mathbb{Z}_m} \sum_{k\in \mathbb{Z}_m} (-1)^{c+k}
\sum_{a=0}^{n-1}  \w^{r^{a+cn} + r^{a+t+kn}}.\end{multline*} We
now make the change of variables: $c = d+k \ (\text{mod} \ m)$
which implies, since $2|m$, that $c = d+k \ (\text{mod} \ 2)$ and
hence $d
 =c-k
=c+k \ (\text{mod} \ 2)$.  As the pair $(c,k)$ varies over all of
$(\mathbb{Z}_m \times \mathbb{Z}_m)$, so does the pair $(d,k)$. We
thus have
\begin{multline*}
\sum_{c\in \mathbb{Z}_m} \sum_{k\in \mathbb{Z}_m} (-1)^{c+k}
\sum_{a=0}^{n-1}  \w^{r^{a+cn} + r^{a+t+kn}} \\ = \sum_{d\in
\mathbb{Z}_m} (-1)^{d} \sum_{a=0}^{n-1} \sum_{k\in \mathbb{Z}_m}
\w^{r^{a+(d+k)n} + r^{a+t+kn}} \\ = \sum_{d\in \mathbb{Z}_m}
(-1)^{d} \sum_{a=0}^{n-1} \sum_{k\in \mathbb{Z}_m}  \w^{(r^{nd} +
r^{t})(r^{a+nk})} \\= \sum_{d=1}^\a (-1)^d \sum_{a=0}^{n-1}
\sum_{k=1}^{\a} \w^{(r^{nd}+r^t)r^{a+nk}}.\end{multline*}\epf

\section{Proof of Approximate Universality and Information Losslessness of Perfect Codes
(Theorem \ref{thm:approximate universality and information
losslessness}) \label{sec:appendix_Proof_ApproxUniversal And
Information Lossless Perfect Codes}} The approximate universality
part of the proof, is based on the derivation of the approximate
universality conditions in \cite{TavVisUniversal_2005}.  It is
reproduced here for
completeness.\\
\begin{proof}
Let $\lambda_1 \leq \lambda_2 \leq \cdots \leq \lambda_n$ and $l_1
\geq l_2 \geq \cdots \geq l_n$ be the ordered eigenvalues of $
H^{\dag}H$ and $\Delta X \Delta X^\dag$ respectively. Irrespective
of the statistics of the channel, in the high-SNR regime, the
probability of no-outage at multiplexing gain $r$, is shown in
\cite{ZheTse} to satisfy\\ $Pr (\text{no-outage}) = Pr \left\{
\sum_{i=1}^{n'} \ln (1+\mbox{SNR} \lambda_i) > \ln( \mbox{SNR}^r)
\right\}$, where $n':=\min(n,n_r)$. Through the Lagrange
multiplier technique we determine $${ d_{E,\text{worst}}^2 =
\inf\limits_{H \notin \text{outage}} \sum_{i=1}^{n'} l_i
\lambda_i}$$ by writing the functional as \
$$J(\lambda_1,\hdots,\lambda_{n'}) = \sum_{i=1}^{n'} l_i \lambda_i
+ \mu \sum_{i=1}^{n'} \ln (1+ \mbox{SNR} \lambda_i) -\mu  r \ln
\mbox{SNR}$$  \ and differentiating w.r.t. $\lambda_i$, we obtain
$ \lambda_i = (\mu/l_i - \mbox{SNR}^{-1}) $.  We then use the
Kuhn-Tucker conditions to verify that the solution ${\lambda_i = (
\mu/l_i - \mbox{SNR}^{-1})^+}$ is what gives the worst possible
$d_{E,\text{worst}}^2$, for $\mu$ such that $${\sum_{i=1}^{n'} \ln
(1+ \mbox{SNR} ( \mu/l_i - \mbox{SNR}^{-1})^+ ) = r \ln
\mbox{SNR}}.$$  Solving the above, we obtain that \vspace{-1pt}
\[ \mu = \overbrace{\mbox{SNR}^{-\left( 1-
\frac{r}{n'}\right)}}^{\psi} \overbrace{ \prod_{i=1}^{n'}
l_i^{\frac{1}{n}}}^{G} \ \ \ \text{and thus} \ \ \ \lambda_i =
\frac{\psi G}{l_i} - \frac{1}{\mbox{SNR}}.\vspace{-3pt}
\]
Substituting this value of $\lambda_i$ in $d_{E,\text{worst}}^2$
and setting $d_{E,\text{worst}}^2
> \mbox{SNR}^\epsilon$ for some $\epsilon>0$, we obtain a condition
on the smallest $n'$ eigenvalues of the code $ \prod_{i=1}^{n'}l_i
> \mbox{SNR}^{n'-r}$, a condition satisfied by CDA
codes with non-vanishing determinant. \cite{isit05_explicit}.

Now moving to perfect codes, we follow the approach in
\cite{HassibiLDCs} to show that the linear dispersion matrices are
unitary.  This property, together with the full rate condition,
establish the information losslessness and the entire theorem.

As the code maps $n^2$ information elements, we consider $n$
linear dispersion matrices $\{A_u\}_{u=1}^n$ of dimension
$n^2\times n$. Starting with
$$ A_{n,i} = diag \left(
G{(i)}\right)_{n\times n}, \ i=1,\cdots,n $$ where $G{(i)}$
represents the $i^{\text{th}}$ row of $G$ in (\ref{eq:G_Matrix}),
we recursively create $$A_{u,i} = \Gamma^{n-u} A_{n,i} , \ \
u=1,2,\cdots,n.$$ Finally $A_u$ is constructed as
 \begin{eqnarray*}
A_u & = &
\begin{array}{|c|} A_{u,0}\\ \vdots \\
A_{u,n-1}\end{array}\end{eqnarray*} It is easy to see that the
unitary nature of $\Gamma$ and $G$ makes each of the $A_u$ unitary
\begin{equation*}\label{eq:rectangular LD matrices} A_u^\dag A_u =
I_n.\end{equation*}
\end{proof}

\begin{center}
ACKNOWLEDGEMENT
\end{center}

The authors would like to thank Saurabha Tavildar and Pramod
Viswanath for making available preprints of their recent work.

\bibliographystyle{IEEEbib}

\end{document}